\documentclass[10pt, twocolumn]{IEEEtran}
\IEEEoverridecommandlockouts  
\usepackage[utf8]{inputenc} 
\usepackage[T1]{fontenc}  
\usepackage{%
algorithm,%
algpseudocode,%
amsfonts,%
amsmath,%
amssymb,%
amsthm,%
bbm,%
caption,%
cite,%
comment,%
extarrows,%
float,%
graphicx,%
hyperref,%
paralist,%
pgfplots,%
subcaption,%
tikz,%
url,%
xcolor,%
}
\usepackage{listings}
\usepackage{indentfirst}

\usepackage{epstopdf}

\usepackage{todonotes}

\pgfplotsset{plot coordinates/math parser=false}
\usepgflibrary{%
patterns,%
}

\usetikzlibrary{%
arrows,%
automata,%
chains,%
decorations.pathreplacing,%
intersections,%
positioning,%
shapes,%
}

\theoremstyle{definition}
\newtheorem{definition}{Definition}
\newtheorem{assumption}{Assumption}
\newtheorem{lemma}{Lemma}
\newtheorem{theorem}{Theorem}

\theoremstyle{remark}

\definecolor{darkgreen}{rgb}{0.0, 0.5, 0.0}

\newcommand{\set}[1]{\left\lbrace#1\right\rbrace}
\newcommand{\setIn}[1]{\mathbbm{1}_{\set{#1}}}

\newcommand{\D}{\mathbb{D}}
\newcommand{\E}{\mathbb{E}}
\newcommand{\N}{\mathbb{N}}
\newcommand{\R}{\mathbb{R}}

\newcommand{\cA}{\mathcal{A}}
\newcommand{\cB}{\mathcal{B}}
\newcommand{\cC}{\mathcal{C}}
\newcommand{\cD}{\mathcal{D}}
\newcommand{\cE}{\mathcal{E}}

\newcommand{\cG}{\mathcal{G}}

\newcommand{\cL}{\mathcal{L}}

\newcommand{\cN}{\mathcal{N}}

\newcommand{\cV}{\mathcal{V}}

\newcommand{\cX}{\mathcal{X}}

\renewcommand{\ge}{\geqslant}
\renewcommand{\le}{\leqslant}

\renewcommand{\ge}{\geqslant}
\renewcommand{\le}{\leqslant}

\begin{document}

\title{\huge Transferable Graphical MARL for Real-Time Estimation in Dynamic Wireless Networks}
	\date{}
	\author{Xingran Chen, \IEEEmembership{}
		Navid NaderiAlizadeh, \IEEEmembership{} 
		Alejandro Ribeiro, \IEEEmembership{} 
		Shirin Saeedi Bidokhti \IEEEmembership{}
		
		\IEEEcompsocitemizethanks {\IEEEcompsocthanksitem Xingran Chen is with the Department of Electrical and Computer Engineering,  Rutgers University, Piscataway, NJ 08854.\quad 
			E-mail: xingranc@ieee.org.
			\IEEEcompsocthanksitem Navid NaderiAlizadeh is with the Department of Biostatistics and Bioinformatics, Duke University, NC 27705.\quad 
			E-mail: navid.naderi@duke.edu.
			\IEEEcompsocthanksitem Alejandro Ribeiro and Shirin Saeedi Bidokhti are with the Department of Electrical and Systems Engineering,  University of Pennsylvania, PA, 19104.\quad 
			E-mail: \{aribeiro, saeedi\}@seas.upenn.edu.
		}
	\IEEEcompsocitemizethanks {Parts of this work have been accepted to IEEE ICASSP 2026.}
	}

\maketitle	
    
\begin{abstract}
We study real-time sampling and estimation of autoregressive Markovian sources in decentralized and dynamic multi-hop networks that share similar structures.
Nodes cache neighboring samples and communicate over wireless collision channels. The objective is to minimize the time-average estimation error and/or the age of information under decentralized policies, which we address by developing a unified graphical multi-agent reinforcement learning framework. A key feature of the framework is its transferability, enabled by the fact that the number of trainable parameters is independent of the number of agents, allowing a learned policy to be directly deployed on dynamic yet structurally similar graphs without re-training. Building on this design, we establish rigorous theoretical guarantees on the transferability of the resulting policies. Numerical experiments demonstrate that (i) our method outperforms state-of-the-art baselines on dynamic graphs; (ii) the trained policies transfer well to larger networks, with performance gains increasing with the number of nodes; and (iii) incorporating recurrence is crucial, enhancing resilience to non-stationarity in both independent learning and centralized training with decentralized execution.
\end{abstract}

\begin{IEEEkeywords}
Real-time sampling and estimation, decentralized strategies, graphical multi-agent reinforcement learning, transferability, age of information
\end{IEEEkeywords}
	
\section{Introduction}\label{sec: Introduction}

Remote sensing and estimation of physical processes have attracted growing attention in wireless networks. Accurate and up-to-date knowledge of system states is crucial for applications such as IoT sensing, robot swarm coordination, autonomous vehicle communication, and environmental monitoring \cite{correlatedAoI, ETLBDM2021}. A fundamental challenge is that minimizing real-time estimation error inherently depends on the timeliness of information updates. Yet, timeliness is hard to guarantee in wireless networks due to delays, unreliable links, and time-varying topology caused by node mobility, failures, and varying connectivity.

In this paper, we study the problem of real-time sampling and estimation in \textit{dynamic multi-hop} wireless networks. Each node (e.g., sensor, device, robot) observes a physical process modeled as a Gauss–Markov source \cite{cxr2020estimation, correlatedAoI} and seeks to maintain accurate, fresh estimates of the processes observed by all other nodes. This task is especially critical in IoT systems, where collaboration relies on continuously updated information. Building on the fundamental challenge above, we identify three major challenges:  
\begin{compactenum}[(i)]
\item \textit{Joint timeliness and estimation:} Real-time information must be collected and used immediately for estimation, rather than waiting for full processing and reliable delivery. This coupling between freshness and accuracy requires new co-design strategies. 
\item \textit{Dynamic network topology:} Node mobility, service variations, and failures cause the network topology to evolve over time, leading to constantly changing connectivity. 
\item \textit{Decentralized decision-making under partial observability:} In large-scale networks, centralized operation is infeasible. Nodes must operate autonomously and adaptively to sustain accurate and timely estimation based only on local observations.  
\end{compactenum}

To tackle the first challenge, the metric Age of Information (AoI) was introduced in 2011 \cite{5984917}. It measures the freshness of information at the receiver and has been adopted as a proxy for real-time estimation error \cite{YSunestimation2020, YSYPEUB2020TIT, cxr2020estimation}. AoI has been studied extensively in diverse settings, including point-to-point channels \cite{SKaul2012Age}, single-hop networks with multiple sources \cite{IKEM2021TMC, RDYSKK2019TIT, IKEM2021INFOCOM, cxr2019}, and multi-hop networks \cite{SFAGKDRB2019, BBASSU2019, VTRTEM2022ToN, ETLBDM2021}. In remote estimation, the relationship between AoI and estimation error is closely linked: fresher packets generally lead to lower instantaneous estimation error, while stale packets yield higher error \cite{cxr2020estimation}.  

Motivated by this connection, a line of work has explicitly investigated estimation error through the lens of AoI. In point-to-point channels, \cite{YSYPEUB2017ISIT, YSYPEUB2020TIT} pioneered the use of AoI to minimize estimation error for Gauss–Markov processes via optimal stopping strategies, with extensions in \cite{TZOYS2021ToN, TZOYS2021SPAWC}. The work in \cite{CHTCHW2022ToN} further unified AoI minimization and remote estimation error minimization. Other metrics have also been considered, such as effective age \cite{CKSKGD2018INFOWK} and the age of incorrect information \cite{AMSKMAAE2020ToN}, leading to optimal policies under zero-wait, sample-at-change, and Markov decision process frameworks. In random access networks, \cite{9589639} proposed an optimal one-bit update strategy, while \cite{SSHSMVBSCRM2022CL} approximated estimation error using the age of incorrect information under slotted ALOHA schemes. Our previous work \cite{cxr2020estimation} established an explicit equivalence between AoI and estimation error and proposed threshold-based transmission policies. However, these works mainly focus on one-hop network models, and their applicability to larger and more complex topologies remains unclear.

To tackle the second challenge, optimal transmission policies for freshness and estimation error have been investigated in multi-hop networks. Most prior work, however, focuses on centralized scheduling over fixed graphs \cite{SFAGKDRB2019, BBASSU2019}, with limited attention to decentralized operation. For example, \cite{VTRTEM2022ToN} proposed three policy classes: only one is a simple decentralized strategy (a stationary randomized policy), while the others rely on centralized control. In ad-hoc scenarios without infrastructure, decentralized mechanisms are indispensable. For instance, \cite{ETLBDM2021} developed a task-agnostic, low-latency method for data distribution, but it treats all packets as identical, ignoring information content beyond freshness. Although recent studies have incorporated network topology into decision-making \cite{VTRTEM2022ToN, NicholasJones2023}, they remain centralized and restricted to fixed graphs. This leaves a critical open problem: designing optimal decentralized mechanisms for dynamic multi-hop networks.

To tackle the third challenge, the complexity and dynamics of network topologies motivate the use of learning-based approaches. Multi-agent reinforcement learning (MARL) has achieved notable success in training multiple agents to coordinate in complex environments \cite{epymarl}. In the real-time sampling and estimation problem, each node can optimize its sampling and transmission policies via MARL. Numerous deep MARL algorithms have been proposed in recent years \cite{epymarl, PobloHernandez2019}, broadly falling into three classes \cite{epymarl}: (a) independent learning, where each node is trained independently \cite{Mingtan1993}; (b) centralized multi-agent policy gradient, e.g., \cite{RyanLowwMAAC2017}, which adopt the centralized training and decentralized execution (CTDE) paradigm; and (c) value decomposition methods, i.e., \cite{PeterSunehag2018}, also using CTDE. MARL has also been applied to wireless domains including resource management \cite{NavidNaderializadeh2021}, power allocation \cite{YasarNasir2019}, edge computing \cite{YutongZhang2021}, and edge caching \cite{NavneetGarg2021}.  

Despite these advances, classical MARL methods, typically parameterized by multi-layer perceptrons, are ill-suited for dynamic graphs because they lack permutation equivariance and transferability: the output may change with node reordering even if the graph is unchanged, and performance degrades further as the topology evolves. This motivates the use of graph neural networks (GNNs) \cite{zhao2023graphbased, FGamaStability2020, LuanaRuiz2020}, which are permutation-equivariant and transferable by design, enabling a graphical MARL framework. To date, the only weakly related prior study is \cite{KunalMenda2019}, which applied deep RL to event-driven multi-agent decision processes. Our framework differs in three key aspects: (a) decision mechanism--our work considers synchronous random access, whereas \cite{KunalMenda2019} focuses on asynchronous policies; (b) decision trigger--our framework is time-driven due to synchronization, while theirs is data-driven; and (c) network topology--our approach explicitly accounts for dynamic topologies, which are not addressed in \cite{KunalMenda2019}. While prior studies have explored related ideas \cite{ETLBDM2021, zhao2023graphbased, KunalMenda2019} and the references therein, an additional advantage of our framework is that it is carefully designed so that the number of trainable parameters is independent of the number of agents. As a result, a learned policy can be directly deployed on dynamically evolving graphs with similar structural characteristics, without requiring re-training. In contrast, existing approaches typically rely on frequent re-training when deployed in real-time decision-making systems, leading to substantial computational and communication overhead.

In this work, we investigate decentralized sampling and remote estimation of $M$ independent Gauss–Markov processes over wireless collision channels in \textit{dynamic yet structurally similar} multi-hop networks. Each node makes real-time decisions on (a) when to sample, (b) whom to transmit to, and (c) what to transmit, with the goal of minimizing the time-average estimation error and/or AoI. As a first theoretical connection, we establish that, when decisions are independent of the Gauss–Markov processes, minimizing estimation error is equivalent to minimizing AoI, which provides a unified problem formulation.  The main contributions of this paper are threefold:
\begin{compactenum}[(i)] 
\item  \textbf{Transferable Graphical MARL Framework with Reduced Learning Cost}. We propose a carefully designed graphical MARL framework that integrates a graphical actor, a graphical critic, and an action distribution operator to jointly determine when to sample, whom to transmit to, and what to transmit in a \textit{decentralized} manner. A key feature of the framework is its \textit{transferability}: because the number of trainable parameters is independent of the number of agents, a learned policy can be directly deployed on dynamically evolving yet structurally similar graphs \textit{without} re-training. Consequently, policies trained on small or moderate networks can be applied to larger-scale graphs, substantially reducing learning costs. This framework further supports the co-design of estimation error and AoI by coupling them within a single  policy.
\item \textbf{New theoretical characterization of transferability}. We establish a rigorous theoretical framework that proves the transferability of the proposed graphical MARL policies across dynamic yet structurally similar networks. In contrast to prior studies on transferability in GNNs, our setting is fundamentally different, as we study a graphical MARL framework in which GNNs serve only as one component. As a result, the proposed theoretical guarantees cannot be directly derived from existing GNN-only analyses. These theoretical results constitute the core novelty of this work.
\item \textbf{Extensive empirical validation}. Experiments demonstrate that (i) the proposed graphical MARL outperforms classical MARL, while the centralized training with decentralized execution (CTDE) mitigates non-stationarity relative to independent learning; (ii) graphical MARL exhibits strong transferability, yielding performance gains in large-scale networks; and (iii) recurrence enhances resilience to non-stationarity in both CTDE and independent learning.
\end{compactenum}
 
\section{System Model}\label{sec:systemModel}

Consider $M$ statistically identical nodes communicating over a {\it connected} undirected graph. Let $\cG = (\cV, \cE)$ denote the graph, where $\cV = \set{1,2,\cdots,M}$ is the node set and $\cE$ is the edge set representing communication links. For each node $i\in\cV$, let $\partial_i = \set{j \mid (i,j)\in\cE}$ denote its neighborhood. Although $\cG$ may evolve over time, we assume that successive graphs remain {\it structurally similar} (Definition~\ref{def: similar graph and signal}). This assumption is broad but practical: for instance, nodes may move continuously while preserving spatial distribution and connectivity rules, or the network size may grow while maintaining a roughly constant average degree. Such situations are common in wireless networks---including mobile ad hoc, vehicular, and UAV/swarm systems---where instantaneous topology changes rapidly but large-scale statistical structure remains stable. For simplicity, we omit the time index $k$ in $\cG$, $\cV$, and $\cE$.

Let time be slotted. Each node $i \in \cV$ observes a physical process $\set{\zeta_{k}^{(i)}}_{k \ge 0}$ in slot $k$, evolving as 
\begin{align}\label{eq: Wiener process}
\zeta_{k+1}^{(i)} = \zeta_{k}^{(i)} + \Lambda_{k}^{(i)},
\end{align}
where $\Lambda_{k}^{(i)}\sim\cN(0,\sigma^2)$ are i.i.d. across all $i$ and $k$ \cite{cxr2020estimation, correlatedAoI}. The processes are mutually independent across nodes, with $\zeta_{0}^{(i)}=0$ for all $i\in\mathcal{V}$.  

Each node is equipped with a cache that stores both its own samples and packets received from neighbors. At the beginning of each slot, a node decides whether to transmit a packet or remain silent. If transmitting, it selects one cached packet (originating from node $\ell$) and one neighbor $j\in\partial_i$ as the receiver. Define $d^{j, \ell}_{i, k}=1$ if node $i$ successfully transmits a packet from node $\ell$ (in its cache) to neighbor $j$ at time $k$, and $d^{j, \ell}_{i, k}=0$ otherwise. By definition, $d^{j, \ell}_{i, k}=0$ for all $k,\ell$ if $j\notin\partial_{i}$.

\begin{figure}
\centering 
\includegraphics[width=0.33\textwidth, height=0.15\textheight]{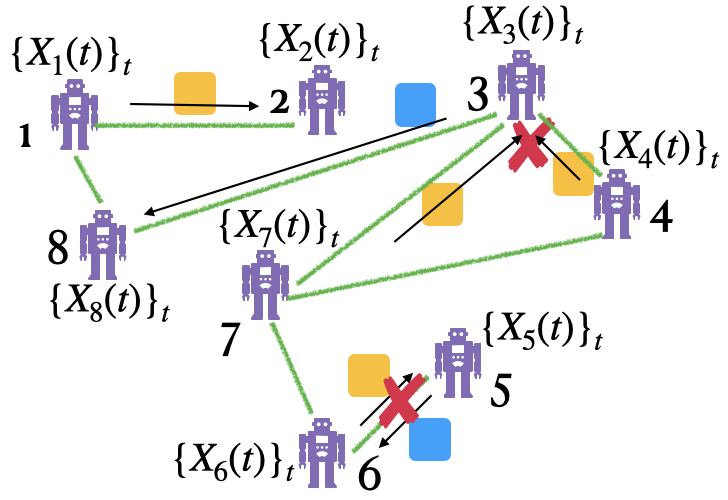} 
\caption{Blue squares indicate packets sampled by the node itself, while yellow squares represent packets received from other nodes. At the shown slot, nodes $1$, $3$, $4$, $5$, $6$, and $7$ attempt transmissions. Collisions occur between nodes $4$ and $7$, and between nodes $5$ and $6$.}
\label{fig_tranmission_process} 
\end{figure}

The communication medium is modeled as a collision channel. If two or more neighboring nodes transmit to the same receiver in the same slot, or if two nodes on opposite ends of an edge transmit to each other simultaneously, all involved transmissions fail. Let $c_{i,k}^j=1$ if a collision occurs on edge $(i,j)$ in the direction $i\to j$ during slot $k$, and $c_{i,k}^j=0$ otherwise. Only the sender observes the collision feedback. Each successful packet delivery takes one slot.  
An illustration of the transmission process is given in Fig.~\ref{fig_tranmission_process}. 

Random access protocols are broadly classified into synchronous (e.g., slotted ALOHA) and asynchronous (e.g., CSMA). In this work we focus on the {\it synchronous} case.

\subsection{Optimization Objectives and  Policies}\label{subsec: Optimization Objectives and  Policies}

Each node $i$ forms an estimate $\hat{\zeta}_{j, k}^{(i)}$ of $\zeta_{k}^{(j)}$ at time $k$. By convention, $\hat{\zeta}_{i, k}^{(i)}=\zeta_{k}^{(i)}$ for all $i, k$, and $\hat{\zeta}_{j, 0}^{(i)}=0$ for all $i, j\in\mathcal{V}$. We use the average sum of estimation errors (ASEE) as the performance metric:
\begin{align}\label{eq: E-allnodeneighbor}
&L^{\pi}=\lim_{K\rightarrow\infty}\E[L_K^{\pi}]\nonumber\\
&L_K^{\pi}=\frac{1}{M^2K}\sum_{k=1}^{K}\sum_{i=1}^{M}\sum_{j=1 }^M\big(\hat{\zeta}_{j,k}^{(i)} - \zeta_{k}^{(j)}\big)^2,
\end{align}
where $\pi\in\Pi$ is a decentralized sampling and transmission policy, and $\Pi$ is the set of all feasible policies. The normalization factor $M^2K$ averages the error over all node pairs and time slots. The optimization problem is then
\begin{equation}\label{eq: optimization}
\min_{\pi\in\Pi}\,\, L^{\pi}.
\end{equation}

Let us refine the notion of policy in \eqref{eq: E-allnodeneighbor} and \eqref{eq: optimization}. 
\begin{definition}\label{def: policy}
A sampling and transmission policy is a sequence of actions 
$\set{\mu_{k}^{(i)}, \nu_{k}^{(i)}}_{i\in\cV, k\ge0}$, where at each slot $k$:  
\begin{compactenum}[(i)]
\item If $\mu_{k}^{(i)}\in\partial_i$, then node~$i$ transmits a cached packet originating from node~$\mu_{k}^{(i)}$ to neighbor $\nu_{k}^{(i)}\in\partial_i$. Here, $\mu_k^{(i)}$ specifies who to communicate with, and $\nu_k^{(i)}$ specifies what to transmit.
\item If $\mu_{k}^{(i)} = i$, then node~$i$ remains silent in slot $k$; conventionally we set $\nu_{k}^{(i)}=i$. This choice implicitly specifies when to sample.
\end{compactenum}
\end{definition}
Compared with classical routing policies, which only determine forwarding paths between sources and destinations, the policies in Definition~\ref{def: policy} are broader: they joint govern sampling, receiver choice, and content selection, allowing each node not only to forward but also to generate and store packets.

Our objective is to design \textit{decentralized} sampling and transmission mechanisms that minimize \eqref{eq: optimization}. At each slot $k$, every node selects its action based only on its local observations and past actions. We distinguish between two classes of policies: \textit{oblivious} and \textit{non-oblivious}. Under oblivious policies, decisions are independent of the underlying processes, and AoI serves as the key decision metric. Under non-oblivious policies, decisions depend on the observed processes themselves, and AoI alone is no longer sufficient.

\subsection{Estimation Error and AoI}\label{subsec: Estimation Error and AoI}

Each node $i$ maintains $M$ {\it virtual queues}. The queue associated with node $j\in\cV$, denoted $Q_{j}^{(i)}$, stores the packets of node $j$ that are cached at node $i$. We assume that each queue $Q_{j}^{(i)}$ has buffer size one: when a new packet from node $j$ arrives, it either replaces the undelivered packet currently in the queue (if any) or is discarded. This assumption is justified by the Markovian nature of the underlying processes--since the most recent packet suffices to characterize the state of the corresponding process, older packets become obsolete.

\begin{figure}
\centering 
\includegraphics[width=0.43\textwidth, height=0.2\textheight]{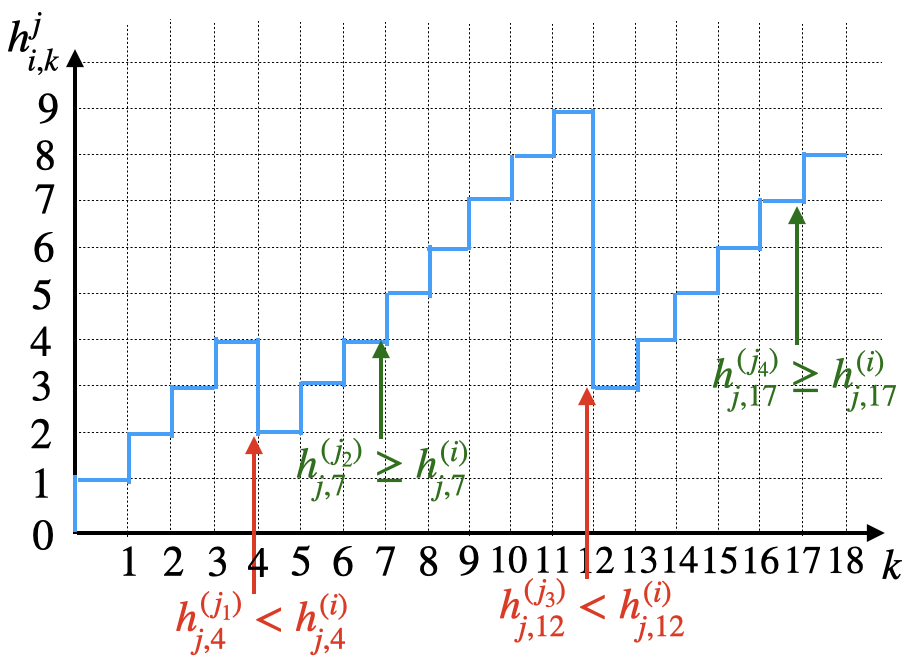} 
\caption{An example trajectory of $h_{j, k}^{(i)}$ is shown: it drops at slots $4$ and $12$ when fresh packets are received from nodes $j_1$ and $j_3$, respectively, and increases at slots 7 and 17 when the received packets from nodes $j_2$ and $j_4$ are stale.}
\label{fig_AoI} 
\end{figure}

Let $\tau_{i, j}$ denote the generation time of the packet stored in $Q_{j}^{(i)}$. The AoI with respect to $Q_{j}^{(i)}$ is defined as \cite{cxr2019}
\begin{align}\label{eq: h_i}
h_{j,k}^{(i)} = k - \tau_{i, j},
\end{align}
with $h_{j,0}^{(i)}=0$ by convention.  When a new packet from node $j$ is delivered to node $i$, it may update $Q_{j}^{(i)}$. When node $i$ receives a packet of node $j$ with generation time $\tau'$, it updates $Q_j^{(i)}$ only if $\tau' > \tau_{i,j}$; otherwise the packet is stale and discarded, since caching it would increase the AoI. Formally, the AoI recursion (as illustrated in Fig.~\ref{fig_AoI}) is
\begin{equation}\label{eq: recursion of h}
h^{(i)}_{j,k+1} =
\begin{cases}
h^{(u)}_{j,k}+1, & \text{if } d_{u,k}^{i,j}=1 \text{ and } h^{(u)}_{j,k}<h^{(i)}_{j,k},\\
h^{(i)}_{j,k}+1, & \text{otherwise}.
\end{cases}
\end{equation}

At the beginning of slot $k$, node $i$ knows the packet currently stored in $Q_{j}^{(i)}$, namely $\zeta_{\tau_{i, j}}^{(j)}$. Its estimate of $\zeta_{k}^{(j)}$ is given by the MMSE (Kalman) estimator, which is optimal in the mean-square sense \cite{correlatedAoI}:
\begin{align}\label{eq: MMSE}
\hat{\zeta}_{j,k}^{(i)} = \mathbb{E}[\zeta_{k}^{(j)}\mid \zeta_{\tau_{i,j}}^{(j)}].
\end{align}
From \eqref{eq: Wiener process} and \eqref{eq: h_i},
\begin{align}\label{eq: recursion of hat}
\zeta_{k}^{(j)} = \zeta_{\tau_{i,j}}^{(j)} + \sum_{\tau=1}^{h_{j,k}^{(i)}}\Lambda_{k-\tau}^{(j)},
\end{align}
and since $\mathbb{E}[\Lambda_{k}^{(j)}]=0$ for all $k$, it follows that
\begin{align}\label{eq: MMSE 1}
\hat{\zeta}_{j,k}^{(i)} = \mathbb{E}[\zeta_{k}^{(j)}\mid \zeta_{\tau_{i,j}}^{(j)}] = \zeta_{\tau_{i,j}}^{(j)}.
\end{align}
The estimate recursion is therefore
\begin{equation}\label{eq: recursion of estimates}
\hat{\zeta}^{(i)}_{j,k+1} =
\begin{cases}
\hat{\zeta}^{(u)}_{j,k}, & d_{u,k}^{i,j}=1 \text{ and } h^{(u)}_{j,k}<h^{(i)}_{j,k},\\
\hat{\zeta}^{(i)}_{j,k}, & \text{otherwise},
\end{cases}
\end{equation}
meaning that node $i$ updates its estimate only when it receives a fresher packet (carrying newer information) from node $u$; otherwise, it keeps its current estimate.

\begin{lemma}\label{lem: Gaussian distribution}
Under oblivious policies, the expected estimation error for process $j$ at node $i$ is proportional to the expected AoI:
\begin{align}\label{eq: Gaussian distribution}
\E\left[\big(\zeta_{k}^{(j)}-\hat{\zeta}_{j,k}^{(i)}\big)^2\right] = \E\left[h_{j,k}^{(i)}\right]\sigma^2.
\end{align}
\end{lemma}
\begin{proof}
At the beginning of slot $k$,
\begin{align*}
\zeta_{k}^{(j)}-\hat{\zeta}_{j,k}^{(i)} = \zeta_{k}^{(j)}-\zeta_{\tau_{i,j}}^{(j)} = \sum_{m=1}^{h_{j,k}^{(i)}} \Lambda_{k-m}^{(j)}.
\end{align*}
Under oblivious policies, $h_{j,k}^{(i)}$ is independent of the Gaussian innovations $\set{\Lambda_{k}^{(j)}}_{j, k}$. Since the $\Lambda_{k}^{(j)}$ are i.i.d. zero-mean with variance $\sigma^2$, Wald’s equality gives
\begin{align*}
\E\left[\zeta_{k}^{(j)}-\hat{\zeta}_{j,k}^{(i)}\right] &= 0,\\
\E\left[\big(\zeta_{k}^{(j)}-\hat{\zeta}_{j,k}^{(i)}\big)^2\right] &= \E[h_{j,k}^{(i)}]\sigma^2.
\end{align*}
\end{proof}

Note that Lemma \ref{lem: Gaussian distribution} does not hold for non-oblivious policies. Finding $\E\left[(\zeta_{k}^{(j)}-\hat{\zeta}_{j, k}^{(i)})^2\right]$ in closed-form is non-trivial and its numerical computation can be intractable when $M$ is large. The challenge arises because even though the estimation error is the sum of $h_{j, k}^{(i)}$ Gaussian noise variables, once we condition on $h_{j, k}^{(i)}$, their distributions change because $h_{j, k}^{(i)}$ can be dependent on the process being monitored. Importantly, Lemma~\ref{lem: Gaussian distribution} implies that, \textit{in the class of oblivious policies}, minimizing the ASEE in \eqref{eq: optimization} is equivalent to minimizing the time-average AoI, i.e., 
\begin{equation}\label{eq: aoioptimization}
\min_{\pi\in\Pi'}\,\, J^{\pi},
\end{equation}
where
\begin{align*}
J^{\pi}=\lim_{K\rightarrow\infty}\mathbb{E}[J_K^{\pi}],\,\,J_K^{\pi}=\frac{1}{M^2K}\sum_{k=1}^{K}\sum_{i=1}^{M}\sum_{j=1 }^M h_{j, k}^{(i)},
\end{align*}
and $\Pi'$ denotes the class of all oblivious policies. Later, we will develop a {\it unified} approach to address both \eqref{eq: optimization} and \eqref{eq: aoioptimization}, as outlined in Section~\ref{sec: GNNRL} onwards.

\section{Preliminaries}\label{sec:Preliminaries}

\subsection{Dec-POMDP and Reinforcement Learning}\label{subsec: Multi-agent Markov Decision Process}
We begin by defining a Decentralized Partially Observable Markov Decision Process (Dec-POMDP) \cite{AmalFeriani2021}. A Dec-POMDP extends a standard Markov Decision Process by incorporating local observations: the true system state is hidden, and each agent only observes a partial, noisy view of it.

\begin{definition}\label{def: MAMDP}
A Dec-POMDP is described by the tuple $\langle M, \cV, S, \set{A_i}_{i\in\cV}, P_s, R, \set{O_i}_{i\in\cV}, P_o, \gamma \rangle$, where
\begin{compactenum}[(i)]
\item $M$: number of agents, with $\cV=\set{1,\dots,M}$. 
\item $S$: set of global states. 
\item $A_i$: action space of agent $i$. 
\item $P_s(s'|s,a)$: state transition probability from $s\in S$ to $s'\in S$ given joint action $a\in\prod_{i\in\cV}A_i$.
\item $R(s,a)$: global reward function shared by all agents. 
\item $O_i$: observation space of agent $i$. 
\item $P_o(o|s)$: observation probability of joint observation $o=(o_1,\dots,o_M)$ in state $s\in S$. 
\item $\gamma\in[0,1]$: discount factor. 
\end{compactenum}
\end{definition}

At each time step $k$, the environment is in state $s_k\in S$ (unobserved by the agents). Each agent $i$ receives a local observation $o_{i,k}\in O_i$, chooses an action $a_{i,k}\in A_i$, and together they form the joint action $a_k=(a_{1,k},\dots,a_{M,k})$. The environment then transitions to $s_{k+1}\sim P_s(\cdot|s_k,a_k)$ and provides a global reward $r_k=R(s_k,a_k)$. The objective is to find a joint policy $\pi$ that maximizes the expected cumulative discounted reward:
\begin{align*}
\max_\pi \;\; \E\left[\sum_{k=0}^\infty \gamma^k r_k\right].
\end{align*}

\subsection{Graph Recurrent Neural Networks}\label{subsec: Graph Recurrent Neural Networks (GRNNs)}

We begin with the classical recurrent neural networks (RNN) formulation \cite{BookGRNN}:
\begin{align}
z_t =& \rho_1(Bx_t + Cz_{t-1}),\,\, 1\le t\le T,\nonumber\\
\hat{y} =& \rho_2(Dz_T), \label{eq: RNN-1}
\end{align}
where the input sequence $\set{x_t}_{t=1}^T$ with $x_t\in\R^n$ is mapped to hidden states $z_t\in\R^n$, and the output $\hat{y}\in\R^{n'}$ estimates the label $y\in\R^{n'}$. Here $B,C\in\R^{n\times n}$ and $D\in\R^{n\times n'}$ are learnable matrices, while $\rho_1,\rho_2$ are pointwise nonlinearities. Given a training set $\set{\set{x_t}_{t=1}^T,y}$, the parameters are obtained by minimizing a loss $\cL(\hat{y},y)$.

Extending to graphs, let $\Xi$ denote a graph shift operator. Replacing matrix multiplications with graph convolutions yields the graph RNN (GRNN):
\begin{align}
z_t =& \rho_1\left(B(\Xi)x_t + C(\Xi)z_{t-1}\right),\,\, 1\le t\le T,\nonumber\\
\hat{y} =& \rho_2\!\left(D(\Xi)z_T\right), \label{eq:GRNN-1}
\end{align}
where $B(\Xi),C(\Xi),D(\Xi)$ are graph convolution filters \cite{LuanaRuiz2020,FernandoGama2019,JianDu2018}. A graph filter such as $B(\Xi)$ is expressed as
\begin{align}\label{eq:graph-filter}
B(\Xi)x = \sum_{k=0}^{K-1} b_k \Xi^k x,
\end{align}
with coefficients $\set{b_k}_k$ and filter order $K$. Analogously, $C(\Xi)$ and $D(\Xi)$ are defined with coefficients $\set{c_k}_k$ and $\set{d_k}_k$.

In practice, each node typically carries multiple features. Collecting them gives the feature matrix $X\in\R^{M\times F}$ for $M$ nodes and $F$ features. Each column corresponds to a graph signal across the network. The GRNN update generalizes to
\begin{align}
Z_t &= \rho_1\left(\cB(\Xi)X_t + \cC(\Xi)Z_{t-1}\right),\,\, 1\le t\le T,\nonumber \\
\hat{Y} &= \rho_2\left(\cD(\Xi)Z_T\right), \label{eq:extend-GRNN-2}
\end{align}
where $Z_t\in\R^{M\times H}$ are hidden states, $\hat{Y}\in\R^{H\times G}$ is the output, and $\cB(\Xi),\cC(\Xi),\cD(\Xi)$ are multi-feature graph convolutions, e.g.,
\begin{align}\label{eq:graph-operators}
\cB(\Xi)X = \sum_{\tau=0}^{K-1} \Xi^\tau X B_k,\quad B_k\in\R^{F\times H}.
\end{align}
Analogously, $\cC(\Xi)$ and $\cD(\Xi)$ are defined with matrices $C_k\in\R^{H\times H}$ and $D_k\in\R^{H\times G}$. For clarity, the GRNN defined in \eqref{eq:extend-GRNN-2} can be compactly expressed as
\begin{align}\label{eq:compact-GRNN}
\hat{Y} = \Phi\left(\cB,\cC,\cD; \Xi, \{X_t\}_{t=1}^T\right).
\end{align}
Notably, the GRNN in \eqref{eq:compact-GRNN} corresponds to a single-layer GRNN block. The extension to an $L$-layer stacked GRNN is straightforward: each layer $\ell$ has its own parameters $\cB^{(\ell)},\cC^{(\ell)},\cD^{(\ell)}$, and the hidden states are recursively updated across layers.

\subsection{Graphons}\label{subsec: Graphons}

We adopt the notion of graphons from \cite{Transferability}.  A \textit{graphon} is a bounded, measurable, and symmetric function $W:[0,1]^2\to[0,1]$, which arises as the limit object of a sequence of dense undirected graphs. A \textit{graphon signal} is a function $X\in L^2([0,1])$.  Intuitively, $(W,X)$ can serve as generative models for graphs and graph signals.

Given $(W,X)$, an $m$-node graph–signal pair $(\Xi_m, x_m)$ is obtained as follows: a point $u_i\in[0, 1]$ is chosen to be the label of node $i$ with $i\in[m]$. For $1\leq i, j\leq m$,
\begin{align}
[\Xi_m]_{ij}&= \text{Bernoulli}\big(W(u_i, u_j)\big)\label{eq: WtoXi}\\
[x_m]_i&=X(u_i)\label{eq: graphonsignaltographsignal}.
\end{align}  
For example, {\it stochastic graphs} can be constructed using the following rule: Let $\{u_i\}_{i=1}^{m}$ be $n$ points sampled independently and uniformly at random from $[0, 1]$. The $m$-node stochastic graph $\cG_{m}$, with graph shift operator $\Xi_{m}$, is obtained from $W$ by \eqref{eq: WtoXi}. 

\begin{definition}[Similar graphs and signals]\label{def: similar graph and signal}
Let $(\Xi_{m_1},x_{m_1})$ and $(\Xi_{m_2},x_{m_2})$ be generated from the same graphon–signal pair $(W,X)$ via the rule \eqref{eq: WtoXi} and \eqref{eq: graphonsignaltographsignal}. 
Then $\Xi_{m_1},\Xi_{m_2}$ are called \emph{similar graphs}, and $x_{m_1},x_{m_2}$ are called \emph{similar signals}.
\end{definition}

Conversely, one can induce graphons from finite graphs. Suppose $\cG_m$ is a graph with graph shift operator $\Xi_m$ and node labels $\set{u_i}_{i=1}^m\subset[0,1]$, and let $x_m$ be a graph signal. 
Define intervals $I_i=[u_i,u_{i+1})$ for $1\le i<m$ and $I_m=[u_m,1]\cup[0,u_1)$, forming a partition of $[0,1]$.  The induced graphon–signal pair $(W_{\Xi_m},X_m)$ is
\begin{align}
W_{\Xi_m}(u,v) &= \sum_{i=1}^m\sum_{j=1}^m [\Xi_m]_{ij}\,\mathbf{1}_{\{u\in I_i\}}\mathbf{1}_{\{v\in I_j\}}, \label{eq: induced graphon}\\
X_m(u) &= \sum_{i=1}^m [x_m]_i\,\mathbf{1}_{\{u\in I_i\}}. \label{eq: graphon signal induced}
\end{align}
Here $W_{\Xi_m}$ is a step-function approximation of the original graphon, while $X_m$ extends node-level signals to the unit interval.

\subsection{Graphon Recurrent Neural Networks}\label{subsec: WNN}

Consider a graphon $W$ and a graphon signal $X\in L^2([0, 1])$. The associated diffusion operator is
\begin{align}\label{eq: original operator}
(T_WX)(v) = \int_{0}^{1} W(u,v)X(u)\,du.
\end{align}
A \textit{graphon filter} is a linear operator $T_{B,W}: L^2([0,1])\to L^2([0,1])$ defined recursively as
\begin{align}\label{eq: diffusion operator}
&(T_{B,W}X)(v) = \sum_{k=0}^{K-1} b_k (T_W^{(k)}X)(v),\nonumber\\ 
&T_W^{(k)}=T_W\circ T_W^{(k-1)},\;\;T_W^{(0)} = I,
\end{align}
with coefficients $\{b_k\}_{k=0}^{K-1}$. Analogously, $T_{C, W}$ and $T_{D, W}$ are defined with coefficients $\set{c_k}_k$ and $\set{d_k}_k$.
Similar to \eqref{eq:extend-GRNN-2}, a \textit{graphon recurrent neural network (WRNN)} is defined as: for $1\le t\le T$,
\begin{align}
Z_t &= \rho_1\left(T_{B,W}X_t + T_{C,W}Z_{t-1}\right),\nonumber \\
\hat{Y} &= \rho_2\left(T_{D,W}Z_T\right).\label{eq:extend-WRNN-2}
\end{align}

To incorporate multiple features, suppose the input at time $t$ is
\begin{align*}
X_t = \set{X_t^f}_{f=1}^{F},\quad X_t^f\in L^2([0, 1]).
\end{align*}
A multi-feature graphon filter maps $F$ input channels to $G$ output channels as 
\begin{align}\label{eq:multi-feature graphon filter}
\left(T_{\cB, W}X_t\right)^g(v) = \sum_{f=1}^{F}\sum_{k=0}^{K-1}[B_k]_{fg}(T_{W}^{(k)}X_t^f)(v),
\end{align}
where $B_k\in\R^{F\times G}$ are learnable matrices. Analogously, $T_{\cC, W}$ and $T_{\cD, W}$ are defined with matrices $C_k\in\R^{H\times H}$ and $D_k\in\R^{H\times G}$.
Given the input sequence $\set{X_t}_{t=1}^T$, the hidden states and output evolve as: for $1\le t\le T$,
\begin{align}
&Z_t = \rho_1\left(T_{\cB,W}X_t + T_{\cC,W}Z_{t-1}\right),\nonumber \\
&\hat{Y} = \rho_2\left(T_{\cD,W}Z_T\right), \label{eq:extend-WRNN-2}
\end{align}
which we abbreviate as
\begin{align}\label{eq: compact WRNN-1}
Y = \Psi\left(T_{\cB},T_{\cC}, T_{\cD}; W, \{X_t\}_{t=1}^T\right).
\end{align}
The WRNN in \eqref{eq: compact WRNN-1} corresponds to a single-layer WRNN block. One can extend it to an $L$-layer stacked WRNN is straightforwardly.

\section{Proposed Graphical MARL Framework}\label{sec: GNNRL}

\begin{figure}
\centering 
\includegraphics[width=0.48\textwidth]{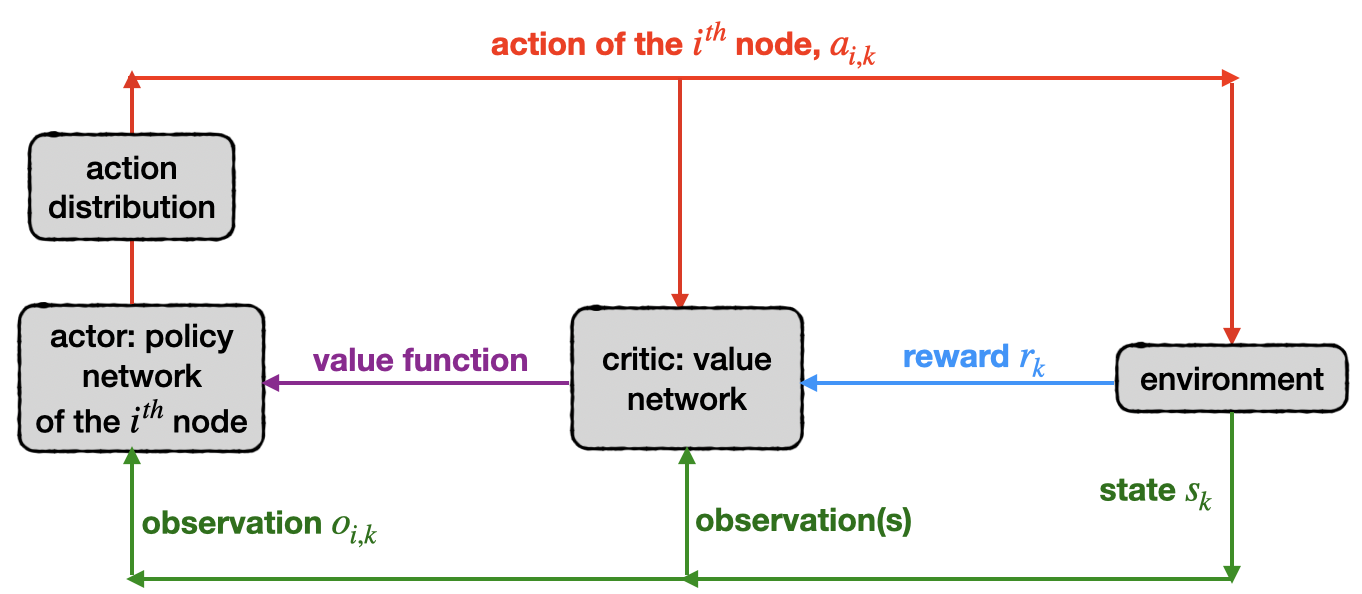} 
\caption{The proposed graphical reinforcement learning framework.}
\label{fig_outline} 
\end{figure}

Fig.~\ref{fig_outline} illustrates our framework, which integrates Dec-POMDP modeling, a GRNN-based actor, a GNN-based critic, and an action distribution operator. We first give an overview, then discuss two key components in detail: (i) graphical actors and critics, and (ii) the action distribution (Sections~\ref{sec: Graph Actor and Critic}--\ref{sec: Action Distribution}). A central contribution of this framework is a scale-invariant parameterization, under which the number of trainable parameters is independent of the number of agents (network size). This enables direct transfer of learned policies across networks of different scales, without retraining or architectural modification.

\subsection{Framework}\label{subsec:MRALframework}

\subsubsection{State and Observations}
Let $\Xi$ denote the adjacency matrix (or graph shift operator) of the network $\cG$. Define $q_{i,k}^j$ as an indicator variable, where $q_{i,k}^j=1$ if  node $i$ sends a packet to node $j$ during time slot $k$, while $q_{i, k}^j=0$ indicates otherwise. At time slot $k$, the environment state $s_k$ includes all processes, estimates, AoIs, communication outcomes, and the current adjacency matrix $\Xi$,
\begin{align}\label{eq: enrivonment}
s_k = \set{\set{\zeta_{k}^{(i)}, \hat{\zeta}_{j, k}^{(i)}, h_{j, k}^{(i)},  c_{i, k}^j, q_{i, k}^j, d_{i, k}^{j, \ell}}_{i, j, \ell\in\cV}, \Xi}.
\end{align}
Node $i$ only observes its local process, cached estimates, local AoIs, collision feedback, past transmissions, and neighborhood $\partial_i$, denoted by $o_{i,k}$,
\begin{align}\label{eq: observations}
o_{i, k} = \set{\set{\zeta_{k}^{(i)}, \hat{\zeta}_{j, k}^{(i)}, h_{j, k}^{(i)},  c_{i, k}^j, q_{i, k}^j, d_{i, k}^{j, \ell}}_{j, \ell\in\cV}, \partial_i}.
\end{align}

\subsubsection{Nodes' Actions}
Given $o_{i,k}$, node $i$ selects an action $a_{i,k}=(\mu_k^{(i)},\nu_k^{(i)})$ using an actor $\pi(o_{i,k};\theta_i)$ and an action distribution $\cA(\cdot;\vartheta_i)$.  
We adopt \textit{parameter sharing} across homogeneous nodes as in prior MARL work~\cite{TabishRashid2018}, such that $\pi(o_{i,k};\theta_i)$ simplifies to $\pi(o_{i,k};\theta)$ and 
$\cA(\cdot \,;\vartheta_i)$ simplifies to $\cA(\cdot \,;\vartheta)$, where $\theta_1=\cdots=\theta_M=\theta$ and $\vartheta_1=\cdots=\vartheta_M=\vartheta$. 
Actions follow
\begin{align*}
a_{i,k} \sim \cA(\pi(o_{i,k};\theta);\vartheta),
\end{align*}
where $(\mu_k^{(i)},\nu_k^{(i)})$ is as in Definition~\ref{def: policy}. In our framework, the actor $\pi(\cdot;\theta)$ is a GRNN. That is, it exactly matches the GRNN mapping $\Phi(\cdot)$ defined in \eqref{eq:compact-GRNN}.

\subsubsection{Rewards}
The per-slot reward is the negative average estimation error
\begin{align*}
r_k = -\frac{1}{M^2}\sum_{i,j\in\cV}(\zeta_k^{(j)}-\hat{\zeta}_{j,k}^{(i)})^2.
\end{align*}
This reward corresponds to the average estimation error across all nodes, motivated by two factors: (i) all nodes are statistically identical, and (ii) they cooperate to minimize the ASEE.  
For oblivious policies, Lemma~\ref{lem: Gaussian distribution} reduces this to the average AoI,
\begin{align*}
r_k = - \frac{1}{M^2}\sum_{i, j\in\cV} h_{j,k}^{(i)}.
\end{align*}
The return from time step $k$ is defined as $\sum_{\tau=0}^{\infty}\gamma^{\tau}r_{k+\tau}$,
which represents the total accumulated discounted reward with discount factor $\gamma$.

\subsubsection{Updating Process}
To evaluate the effectiveness of actions, we introduce a value network (critic). The critic estimates the expected return given the current state, and its 
feedback is used to update the actor. During training, the critic updates its parameters by minimizing the discrepancy between the estimated and realized returns. 
Formally, the critic is parameterized as $\pi'(\cdot\,;\phi)$,
where $\phi$ are its learnable parameters under parameter sharing, the critic employs a simpler GNN architecture (a recurrence-free version of 
\eqref{eq:compact-GRNN} with $T=1$) for computational efficiency.

We adopt two well-known advantage actor-critic variants: IPPO and MAPPO~\cite{epymarl, Volodymyr2016}.%
\footnote{Our framework builds on Proximal Policy Optimization (PPO), a popular on-policy reinforcement learning algorithm. 
Recent work has shown that MAPPO, a PPO-based method, is competitive with or outperforms common off-policy approaches such as MADDPG, QMix, and RODE in terms of both sample efficiency and wall-clock time \cite{computationexplexity}.} IPPO corresponds to independent learning with decentralized actor-critic pairs.  In contrast, MAPPO employs a CTDE, using a centralized critic and decentralized actors.

\subsection{Graphical Actor and Critic}\label{sec: Graph Actor and Critic}

We now present the construction of the actor and critic. Since nodes in a wireless network naturally form a graph topology, a graph-based learning model is appropriate. Conventional neural networks are not permutation-invariant--reordering nodes may change the output even if the environment is unchanged--so we employ GNNs, which are inherently permutation-invariant~\cite{Transferability}.

\subsubsection{Actor}
For the actor, we adopt GRNNs, which jointly exploit the graph structure and temporal dynamics. GRNNs are provably permutation-equivariant and stable to graph perturbations \cite[Proposition~1]{Transferability}, and have been shown to outperform both GNNs and RNNs.

The local graph representation and associated features for node $i$ are defined as follows:

\begin{compactenum}[(i)]
\item \textit{Local graph.} Let $\cG^{(i)}$ denote the local graph centered at node $i$, with adjacency matrix $\Xi^{(i)}\in\R^{M\times M}$ that retains only the edges between node $i$ and its neighbors:
\begin{align*}
[\Xi^{(i)}]_{uv} =
\begin{cases}
1, & u=i, v\in\partial_i \text{ or } v=i, u\in\partial_i,\\
0, & \text{otherwise}.
\end{cases}
\end{align*}

\item \textit{Node features.} Each node $j$ in $\cG^{(i)}$ is assigned a feature vector
\begin{align}\label{eq: nodefeatureerror}
v_{j, k}^{(i)} = [(\zeta_{k}^{(j)}-\hat{\zeta}_{j, k}^{(i)})^2,\, h_{j, k}^{(i)},\, c_{i,k}^j,\, q_{i, k}^j,\, [\Xi^{(i)}]_{ji}].
\end{align}
For oblivious policies (Lemma~\ref{lem: Gaussian distribution}), this reduces to
\begin{align}\label{eq: nodefeatureerror0}
v_{j, k}^{(i)} = [h_{j, k}^{(i)},\, c_{i,k}^j,\, q_{i, k}^j,\, [\Xi^{(i)}]_{ji}].
\end{align}
Here edge features are not included explicitly, since connectivity is encoded in $\Xi^{(i)}$.

\item \textit{Graph feature matrix.} Let $F$ denote the feature dimension. According to \eqref{eq: nodefeatureerror} and \eqref{eq: nodefeatureerror0}, $F=5$ for non-oblivious and $F=4$ for oblivious policies. 
The node features $\{v_{j,k}^{(i)}\}_j$ are stacked into a feature matrix $x_{0}^{(i)}\in\R^{M\times F}$.

\item \textit{GRNN actor.} The output of the actor is 
\begin{align}\label{eq: compact GRNN-V}
\hat{y}^{(i)} = \Phi\left(\cB,\cC,\cD; \Xi^{(i)}, \set{x_0^{(i)}}_{t=1}^T\right),
\end{align}
where $\Phi$ is the GRNN operator defined in \eqref{eq:compact-GRNN}. Here, we slightly abuse the notation: $\set{x_0^{(i)}}_{t=1}^{T}$ represents the sequence of input feature matrices over the past $T$ time slots, which are fed into the GRNN to capture temporal dependencies. The policy network parameters are collected as $\theta=\set{\cB,\cC,\cD}$.

\end{compactenum}

\noindent Thus, from \eqref{eq: compact GRNN-V}, the resulting actor can be compactly expressed as
\begin{align*}
\pi(o_{i,k};\theta) = \hat{y}^{(i)}\in\R^{M\times G}.
\end{align*}
The final representation $\hat{y}^{(i)}$ is then mapped into an action distribution 
(see Section~\ref{sec: Action Distribution}) from which $(\mu_k^{(i)},\nu_k^{(i)})$ is sampled.

\subsubsection{Critic}
For the critic, we use a simpler recurrence-free GNN, which provides stable value estimation and is more computationally efficient.

\begin{compactenum}[(i)]
\item \textit{IPPO.} Each node $i$ maintains its own critic with the same structure as \eqref{eq: compact GRNN-V}, but without recurrence ($T=1$). Thus, the critic is implemented as a GNN:
\begin{align*}
\pi'_i(o_{i,k};\phi) = \tilde{y}^{(i)} = \Phi\left(\cB,\cC,\cD; \Xi^{(i)}, x_0^{(i)}\right)\in\R^{M\times G}.
\end{align*} 

\item \textit{MAPPO.} A centralized critic has access to the global state $s_k$. We represent its input as a complete graph: each node $i$ is assigned a feature $|\partial_i|$ (node degree), while each directed edge $i\to j$ is assigned features defined in \eqref{eq: nodefeatureerror} for non-oblivious policies and \eqref{eq: nodefeatureerror0} for oblivious policies. For pairs $(i,j)$ where $j\notin\partial_i$, we introduce virtual edges so that the critic always receives a fully connected graph.

All node and edge features are stacked into a feature matrix $\tilde{x}_0$, which encodes the entire state $s_k$. The critic is then implemented as
\begin{align*}
\pi'(s_k;\phi) = \tilde{y} = \Phi\left(\cB,\cC,\cD; \Xi', \tilde{x}_0\right)\in\R^{M\times G},
\end{align*}
where $\Xi'$ is the adjacency matrix of the complete graph.
\end{compactenum}

\subsection{Action Distribution}\label{sec: Action Distribution}

The final output of the actor for node $i$, obtained from \eqref{eq: compact GRNN-V}, is denoted by $\hat{y}^{(i)}$. To decide an action $(\mu_{k}^{(i)}, \nu_{k}^{(i)})$, we map $\hat{y}^{(i)}$ into a probability distribution over all feasible (\textit{packet-origin, next-hop}) pairs. Here, $\mu_{k}^{(i)}$ specifies the origin of the cached packet to be forwarded, and $\nu_{k}^{(i)}\in\partial_i$ specifies the neighbor to which this packet will be transmitted. 
This distribution is produced by an \textit{action distribution} operator.
Formally, we define
\begin{align}\label{eq: actionsampling}
\cA(\hat{y}^{(i)};\vartheta) = F_\text{softmax}\left(\hat{y}^{(i)} \,\vartheta\, (\hat{y}^{(i)})^T\right),
\end{align}
where $\vartheta\in\mathbb{R}^{G\times G}$ is a learnable parameter matrix. 
The operator $\cA(\cdot;\vartheta)$ produces a categorical distribution, from which the action  
\begin{align*}
(\mu_{k}^{(i)}, \nu_{k}^{(i)}) \sim \cA(\hat{y}^{(i)};\vartheta)
\end{align*}
is sampled. 

By construction, the number of trainable parameters in both the actor and critic depends only on the feature dimension $F$ and a hyper-parameter $G$ (see \eqref{eq:GRNN-1}, \eqref{eq:graph-filter}), while the parameters in the action distribution operator (see $\vartheta$) depend only on the hyper-parameter $G$. Consequently, the total number of trainable parameters is independent of the number of agents $M$ (network size).

\section{Fundamental Analysis}\label{sec: Permutation Invariance and Transferability}
In this section, we highlight the main advantage of the proposed framework: transferability. Transferability means the framework remains effective when networks evolve according to the similarity rule in Definition~\ref{def: similar graph and signal}, enabling its use on graphs with different numbers of nodes as long as they share structural similarity. In contrast to the conventional view of learning as the search for an optimal representation tied to a specific dataset or network, our focus here is on obtaining a representation that remains \textit{sufficient} for solving the task across evolving networks. This shift of emphasis from optimality to transferability captures the essence of our contribution: the framework is designed not for a single static scenario, but for adaptability across structurally similar systems.

It is worth noting that, in classical reinforcement learning, convergence guarantees exist only under restrictive conditions~\cite{10.5555/2998687.2998733,580874}. When extended to multi-agent environments, establishing convergence becomes even more challenging---indeed, no general proof currently exists, and the problem remains open~\cite{Wong2023DeepMARL,marl-book}. This underscores the importance of frameworks, such as ours, that prioritize transferability over strict convergence guarantees.

\subsection{Transferability in GRNNs}

The class of GNNs constructed using graph filters possesses the property of transferability, which can be rigorously established via graphons. In \cite{Firstgraphon}, the authors first introduced graphons to analyze transferability of graph filters, proving that the output of graph filters converges (in the induction sense) to that of the corresponding graphon filter. Subsequent works \cite{NKABSV2020, Transferability, MMGL2021} demonstrated that graph sequences obtained via sampling procedures can converge to a graphon in the homomorphism density sense.

Since GNNs enjoy transferability, it is natural to expect GRNNs, as their temporal extension, to inherit this property. The key idea, inspired by \cite{Transferability}, is as follows: given any graphon $W$ and a graphon signal $X$, we construct a WRNN. By approximating $(W,X)$ with a finite graph $\Xi_m$ and graph signal $x_m$, we obtain a GRNN. If the output of the GRNN converges to that of the WRNN, then the outputs for two similar graphs with corresponding graph signals must also be close.

Formally, let $W$ be a graphon, $\set{X_t}_{t=1}^T$ a graphon signal, and $\mathcal{G}_m$ an $m$-node graph with node labels $\{u_i\}_{i=1}^m$. 
From $(W,\set{X_t}_{t=1}^T)$ we obtain a finite pair $(\Xi_m,\set{x_{t,m}}_{t=1}^{T})$ via \eqref{eq: WtoXi} and \eqref{eq: graphonsignaltographsignal}; 
conversely, $\left(\Xi_m,\set{x_{t,m}}_{t=1}^{T}\right)$ induces $\left(W_{\Xi_m},\set{X_{t,m}}_{t=1}^{T}\right)$ via \eqref{eq: induced graphon} and \eqref{eq: graphon signal induced}. A graphon filter $T_{B,W}$ is defined in \eqref{eq: diffusion operator}, and $B(\Xi_m)$ denotes a graph filter \eqref{eq:graph-filter} instantiated from $T_{B, W}$ on the graph $\Xi_m$. Based on \eqref{eq: compact WRNN-1}, we denote 
\begin{align}
&Y = \Psi\left(T_{\cB},T_{\cC}, T_{\cD}; W, \{X_t\}_{t=1}^T\right),\label{eq:WRNNY}\\
&Y_m = \Psi\left(T_{\cB},T_{\cC}, T_{\cD}; W_{\Xi_m}, \{X_{t, m}\}_{t=1}^T\right).\label{eq:WRNNYn}
\end{align}

\begin{definition}(See \cite[Definition~4]{Transferability})\label{def: c-band cardinality of W}
The $\epsilon$-band cardinality of a graphon $W$, denoted by $\kappa_W^\epsilon$, is the number of eigenvalues $\lambda_i$ of $T_W$ with absolute value larger or equal to $\epsilon$, i.e., 
\begin{align*}
\kappa_W^\epsilon = \#\{\lambda_i: |\lambda_i|\geq \epsilon\}.
\end{align*}
\end{definition}

\begin{definition}\label{def: c-eigenvalue margin}
(See \cite[Definition~5]{Transferability}) For two graphons $W$ and $W'$, the $\epsilon$-eigenvalue margin, denoted by $\delta_{WW'}^\epsilon$, is given by
\begin{align*}
\delta_{WW'}^\epsilon = \min_{i, j\neq i}\{|\lambda_i(T_{W'}) - \lambda_j(T_W)|: |\lambda_i(T_{W'})|\geq \epsilon\},
\end{align*}
where $\lambda_i(T_{W'})$ and $\lambda_i(T_W)$ denote the eigenvalues of $T_{W'}$ and $T_W$, respectively. 
\end{definition}

\begin{assumption}\label{assu: filter Lipschitz}
(See \cite[Assumption~1]{Transferability}) The spectral response of the convolutional filter of $T_{B, W}$, defined as  $b(\lambda)=\sum_{k=0}^{K-1}b_k\lambda^k$, is $\Omega$-Lipschitz in $[-1, -\epsilon]\cup[\epsilon, 1]$ and $\omega$-Lipschitz in $(-\epsilon, \epsilon)$, with $\omega<\Omega$. Moreover, $|b(\lambda)|<1$.
\end{assumption}

Under Assumption~\ref{assu: filter Lipschitz}, \cite[Theorem1]{Transferability} shows that a graphon filter can be approximated by a graph filter on a large graph sampled from the same graphon. The approximation error is influenced by three factors: (i) the distance between the graph and the graphon, representing the graph sampling error; (ii) the distance between the graphon signal and the graph signal, representing the signal sampling error; and (iii) the parameters $(\epsilon, w)$ in Assumption~\ref{assu: filter Lipschitz}, which relate to the design of the convolutional filter. This implies that, by designing a convolutional filter with smaller  $\epsilon$ or $w$ yield better transferability.

\begin{assumption}\label{assu: activation function normalized Lipschitz}
(See \cite[Assumption~5]{Transferability})  The activation functions are normalized Lipschitz, i.e., $|\rho(x)-\rho(y)|\leq |x-y|$, and $\rho(0)=0$.
\end{assumption}

These definitions and assumptions set the stage for analyzing the transferability of WRNNs. In particular, the approximation error between a WRNN and its GRNN instantiation can be attributed to three sources: (i) graph sampling, (ii) signal sampling, and (iii) filter design. The next theorem formalizes this decomposition.

\begin{theorem}[transferability in GRNNs]\label{thm: WRNN approximation on a generic graph}
Let $Y$ and $Y_{m}$ be defined in \eqref{eq:WRNNY} and \eqref{eq:WRNNYn}, respectively. Suppose the convolutional filters in the WRNN satisfy Assumption~\ref{assu: filter Lipschitz}, and $\rho_1$ and $\rho_2$ satisfy Assumption~\ref{assu: activation function normalized Lipschitz}. Assume the input and output feature dimensions satisfy $F=G=1$, and define
\begin{align*}
\eta_1&=\max_{1\le t\le T}\|X_t\|,\\
\eta_2&=\max_{1\le t\le T}\|X_t-X_{t,m}\|.
\end{align*}
Then, for any $0<\epsilon\leq 1$, it holds that\footnote{In Theorem~\ref{thm: WRNN approximation on a generic graph}, the norm $\|\cdot\|$ represents the output of the graph filters converges in the induction sense to the output of the graphon filter. As noted by \cite{SMRLGK2023}, this norm is $\|\cdot\|_{L^2[0, 1]}$. Mathemtically, $\|\cdot\|_{L^2[0, 1]}$ is defined as $\|f\|_{L^2[0, 1]} = \Big(\int_{0}^{1}|f(x)|^2dx \Big)^{\frac{1}{2}}$. For the remainder of this work, we abbreviate $\|\cdot\|_{L^2[0, 1]}$ as $\|\cdot\|$.}
\begin{align}\label{eq: extend WNN approximation on a generic graph2}
\|Y - Y_m\|\leq\frac{T(1+T)}{2} (\Theta_1+\Theta_3)\eta_1 +T \Theta_2\eta_2.
\end{align}
where $\Theta_1 = (\Omega+\frac{\pi\kappa^\epsilon_{W_{\Xi_m}}}{\delta^\epsilon_{WW_{\Xi_m}}})\|W - W_{\Xi_m}\|$,  $\Theta_2 = \Omega\epsilon + 2$, and $\Theta_3 = 2\omega\epsilon$.
\end{theorem}
\begin{proof}
The proof is given in Appendix~\ref{App: proof of transferability in GRNN}.
\end{proof}

Theorem~\ref{thm: WRNN approximation on a generic graph} demonstrates that the output of a WRNN can be approximated by a GRNN on a large graph sampled from the same graphon. The approximation error consists of three main components:
\begin{compactenum}
\item Graph sampling error: $\frac{T(1+T)}{2}\Theta_1\|X\|$, which decreases as the distance between the sampled graph and the graphon ($\Theta_1$) becomes smaller. 
\item Signal sampling error: $T\Theta_2\|X-X_m\|$, which decreases when the sampled graph signal better approximates the graphon signal.
\item Filter design error: $\frac{T(1+T)}{2} \Theta_3\|X\|$, which can be reduced by choosing convolutional filters with smaller parameters $\epsilon$ or $w$, as described in \cite{Transferability}. Transferability property holds only for convolutional filters built on graph filters \cite{Transferability}, such as GCNConv, TAGConv \cite{GNNsheet}.
\end{compactenum}
The dependence on recurrence depth $T$ reflects natural error accumulation in recurrent architectures: errors introduced at each step propagate forward and affect all subsequent steps. Summing these contributions yields a quadratic factor $\sum_{t=1}^T t=\frac{T(1+T)}{2}$, which explains the $\frac{T(1+T)}{2}$ term in the bound.

\subsection{Transferability in the Action Distribution}\label{subsec:TransferabilityActionDist}
So far, we have shown that GRNNs are transferable. We now turn to the transferability of the action distribution \eqref{eq: actionsampling}. Since action distributions are discrete, we use a graphon-based approach: compare them with the limit action distribution. At each learning step, the matrix $\vartheta\in\R^{G\times G}$ in \eqref{eq: actionsampling} is fixed, though it is updated across steps.

In this subsection, we drop the assumption $G=1$ in Theorem~\ref{thm: WRNN approximation on a generic graph}. To define the limit action distribution, we introduce labels
\begin{align}\label{eq: label F}
f_i = \frac{i-1}{G},\quad 1\leq i\leq G,
\end{align}
and intervals $I_i=[f_i, f_{i+1})$ for $i=1,2,\cdots,G-1$, with $I_{G}=[f_{G}, 1]\cup[0, f_1)$. Using \eqref{eq: induced graphon}, the matrix $\vartheta$ induces a graphon,
\begin{align}
&W_{\vartheta}(u, v) = \sum_{i=1}^{G}\sum_{j=1}^{G}[\vartheta]_{ij}\setIn{u\in I_i}\setIn{v\in I_j}\label{eq: induced graphon F}.
\end{align}

Let $\{X^{(1)}_t\}_{t=1}^{T}$, $\{X^{(2)}_t\}_{t=1}^{T}$ be two sequences of graphon signals, and let $\{X_{t,m}^{(1)}\}_{t=1}^{T}$, $\{X_{t,m}^{(2)}\}_{t=1}^{T}$ be their induced graphon signals. Based on \eqref{eq: compact WRNN-1}, we denote 
\begin{align}
&Y^{(1)} = \Psi\left(T_{\cB},T_{\cC}, T_{\cD}; W, \{X_t^{(1)}\}_{t=1}^T\right),\label{eq:WRNNY1}\\
&Y^{(2)} = \Psi\left(T_{\cB},T_{\cC}, T_{\cD}; W, \{X_t^{(2)}\}_{t=1}^T\right),\label{eq:WRNNY2}\\
&Y_{m}^{(1)} = \Psi\left(T_{\cB},T_{\cC}, T_{\cD}; W_{\Xi_m}, \{X_{t, m}^{(1)}\}_{t=1}^T\right),\label{eq:WRNNYn1}\\
&Y_{m}^{(2)} = \Psi\left(T_{\cB},T_{\cC}, T_{\cD}; W_{\Xi_m}, \{X_{t, m}^{(2)}\}_{t=1}^T\right).\label{eq:WRNNYn2}
\end{align}
Define
\begin{align}\label{eq:Delta}
\eta_3 \triangleq \max_{j\in\set{1,2}} \|Y^{(j)} - Y_m^{(j)}\|.
\end{align}
By Theorem~\ref{thm: WRNN approximation on a generic graph}, the error $\|Y^{(j)} - Y_m^{(j)}\|$ is small for each $j\in\set{1,2}$. Thus, $\eta_3$ is small.

Let $T_{W_{\vartheta}}$ be the operator associated with $W_{\vartheta}$ (see \eqref{eq: original operator}). We define two functionals:
\begin{align}
\cX(Y^{(1)}, Y^{(2)}) &\triangleq \langle Y^{(1)}, T_{W_\vartheta}Y^{(2)}\rangle, \label{eq: extended action distribution1}\\
\cX_m(Y_m^{(1)}, Y_m^{(2)}) &\triangleq \langle Y_m^{(1)}, T_{W_\vartheta}Y_m^{(2)}\rangle, \label{eq: extended action distributionn}
\end{align}
where $\langle\cdot,\cdot\rangle$ denotes the inner product. Analogous to \eqref{eq: actionsampling}, each $Y^{(j)}$ can be viewed as a row of $\hat{y}^{(i)}$, and $\cX(Y^{(1)}, Y^{(2)})$ corresponds to an entry of the matrix $\hat{y}^{(i)}\vartheta \left(\hat{y}^{(i)}\right)^T$. The same interpretation holds for the finite-dimensional case with $Y_m^{(j)}$ and $\cX_n(Y_m^{(1)}, Y_m^{(2)})$.

The continuous version of the softmax in \eqref{eq: actionsampling} is defined for the measurable function $\cX: \R\times\R\to\R$ as: for any $Y^{(1)}, Y^{(2)}\in\R$,
\begin{align}\label{eq: continuous softmax}
\left(\tilde{F}_{\text{softmax}}\cX\right)(Y^{(1)}, Y^{(2)}) \triangleq \frac{e^{\cX(Y^{(1)}, Y^{(2)})}}{\int_{\R}\int_{\R}e^{\cX(y_1, y_2)}dy_1dy_2}.
\end{align}
Thus, the {\it limit action distributions} are defined as 
\begin{align}
\tilde{\cA}& \triangleq \tilde{F}_{\text{softmax}}\cX,\label{eq: extended action distribution}\\
\tilde{\cA}_m& \triangleq \tilde{F}_{\text{softmax}}\cX_m.\label{eq: extended action distributionn}
\end{align}
\begin{theorem}[Transferability in action distributions]\label{thm: action distribution A}  
Let $Y^{(1)}$, $Y^{(2)}$, $Y^{(1)}_m$, and $Y^{(2)}_m$ be defined in \eqref{eq:WRNNY1}--\eqref{eq:WRNNYn2}, respectively. Suppose the convolutional filters in the WRNN satisfy Assumption~\ref{assu: filter Lipschitz},  and let $\rho_1,\rho_2$ be as in Theorem~\ref{thm: WRNN approximation on a generic graph}. Let $\eta_3$ be in \eqref{eq:Delta}. Then for any $0<\epsilon\le 1$,  
\begin{align}\label{eq: action distribution-A}   
&\left|\tilde{\cA}(Y^{(1)}, Y^{(2)}) - \tilde{\cA}_m(Y^{(1)}_m, Y^{(2)}_m)\right| \nonumber \\
&\le \Gamma \|T_{W_\vartheta}\|\left(\|Y^{(2)}\|
+ \|Y_m^{(1)}\|\right)\eta_3,   
\end{align}  
where $\Gamma$ is a constant independent of the WRNN in \eqref{eq: compact WRNN-1}.  
\end{theorem}  	
\begin{proof}
The proof is given in Appendix~\ref{App: action distribution AA}.
\end{proof}

From Theorem~\ref{thm: WRNN approximation on a generic graph}, the output discrepancy $\eta_3$ defined in \eqref{eq:Delta} can be made small by properly designing $T_{\cB, W}$, $T_{\cC, W}$, and $T_{\cD, W}$ in \eqref{eq: compact WRNN-1}. Hence, Theorem~\ref{thm: action distribution A}  implies that, for any given pair of input sequences $\{X_t\}_{t=1}^{T}$ and $\{X_{t,m}\}_{t=1}^{T}$, the \textit{pointwise} distance between the limit action distributions, i.e., $\cA(Y^{(1)}, Y^{(2)})$ and $\cA_n(Y^{(1)}_m, Y^{(2)}_m)$, can be small when $\eta_3\to0$. Consequently, the transferability of action distributions is inherited from the output transferability of the WRNNs, and the total error is again governed by the graph sampling error, the signal sampling error, and the filter design error.

Combining Theorems~\ref{thm: WRNN approximation on a generic graph} and~\ref{thm: action distribution A}, we conclude that the proposed framework is transferable. This transferability holds not only across similar graphs of the same size but also across graphs of different sizes. Practically, training GRNNs on very large networks is challenging because (i) full graph knowledge is often unavailable, and (ii) matrix multiplications become costly as the network size grows. Transferability addresses these issues by allowing models trained on smaller graphs to generalize effectively to larger ones.

Finally, good transferability requires sufficiently large network size, since size (or density) directly controls the error bound. In practice, this condition is easily met in IoT and 6G wireless networks, where the large number of devices naturally preserves transferability.

\section{Experimental Results}\label{sec: Simulations}

We confirm our analysis through numerical simulations. The experimental setup and parameters are detailed in Section~\ref{subsec: Experimental Setup}, baselines are defined in Section~\ref{sec: simulation baseline}, and numerical results and discussions are presented in Section~\ref{sec: Numerical Results}.

\subsection{Experimental Setup}\label{subsec: Experimental Setup}

We evaluate the proposed algorithms on both synthetic and real networks:
\begin{compactenum}
\item Synthetic networks. Two graph families are considered: Watts–Strogatz graphs and stochastic block models, each with $N=10$ nodes.\footnote{Experiments fail if $M\geq12$ due to limited computational (GPU) resources.} For Watts–Strogatz graphs, the rewiring probability is set to $0.5$. In stochastic block models, nodes are divided into two communities with intra- and inter-community connection probabilities of $0.6$ and $0.4$, respectively. These models are widely used in practice: the Watts–Strogatz model is useful for modeling heterogeneous sensor networks \cite{WSpractice}, while the stochastic block model is valuable for community detection in large-scale data networks \cite{SBMpractice}.  
\item Real-world network. We use the {\it aus\_simple} topology from \cite{topologyzoo}, consisting of $7$ connected nodes. 
\end{compactenum}

For both synthetic and real-world networks, each learning episode has a time horizon of $1024$ steps, and training runs for a total of $3000$ episodes. We focus on dynamic scenarios: for the synthetic graphs, a new graph from the same family is sampled at the start of each episode; for the real network, the topology is fixed but node labels are randomly permuted at the beginning of each episode, ensuring variability in node ordering even when the underlying topology does not change\footnote{A fixed graph with permuted node indices is a special case of “similar graphs,” where the topology is unchanged but labels vary.}. In essence, synthetic episodes vary the graph structure within a family, while real-world episodes keep the topology identical and vary only the node labels via random re-numbering.

For evaluation during training, we pause every $10$ episodes and test the current model on a fixed set of $30$ held-out tasks: $30$ test graphs for the synthetic case and $30$ test episodes (distinct random permutations) for the real network. We report the aggregate performance over these $30$ test cases.

Regarding model design, actor networks are implemented with GRNNs and critic networks with GNNs. Unless otherwise noted, both use $L=2$ layers with hidden width $64$, and the number of recurrent rounds is set to $T=2$. Many GNN modules are available~\cite{GNNsheet}; in our experiments, we select suitable modules for the actor and critic. Detailed GRNN/GNN architectural choices are reported in Table~\ref{tableparameters2}, and the remaining model/RL hyperparameters are summarized in Table~\ref{tableparameters3}.

\begin{table}[htbp]
\centering
\begin{tabular}{ |c|c| }
\hline
\textbf{GRNN/GNN architectural parameter} & \textbf{Value} \\
\hline
Number of layers $L$ & $2$ \\
\hline
Hidden width per layer & $64$ \\
\hline
Recurrent rounds $T$ & $2$ \\
\hline
GNN module (actor) & GCNConv \\
\hline
GNN module (critic, IPPO) & TAGConv \\
\hline
GNN module (critic, MAPPO) & GINEConv \\
\hline
\end{tabular}
\caption{GRNN/GNN architectural parameters.}
\label{tableparameters2}
\end{table}

\begin{table}[htbp]
\centering
\begin{tabular}{ |c|c| }
\hline
\textbf{Other parameter} & \textbf{Value} \\
\hline
Variance of $\Lambda_{i,k}$ ($\sigma^2$) & $1$ \\
\hline
WS rewiring probability (synthetic) & $0.5$ \\
\hline
Number of nodes (synthetic) & $10$ \\
\hline
Number of nodes (real) & $7$ \\
\hline
\# test graphs/episodes per evaluation & $30$ \\
\hline
Learning rate & $0.0003$ \\
\hline
Steps per episode & $1024$ \\
\hline
Batch size & $10$ \\
\hline
Discount factor $\gamma$ & $0.99$ \\
\hline
\end{tabular}
\caption{Training and evaluation hyperparameters.}
\label{tableparameters3}
\end{table}

\subsection{Baselines}\label{sec: simulation baseline}

We compare our framework against three baselines: (i) classical MARL policies, (ii) adaptive uniform transmitting policies, and (iii) adaptive age-based policies.
\begin{compactenum}
\item {\it Classical MARL policies}: We use the IPPO and MAPPO implementations from \cite{epymarl}. The key difference between classical MARL and graphical MARL lies in the network architecture: is the network architecture: in the former, the actor and critic are fully connected neural networks or recurrent neural networks, whereas in the latter they are built from graph-convolutional layers and an action distribution.

\item {\it Uniform transmitting policies}: Each node with cached packets transmits to adjacent nodes with equal probability. The degree of node $i$ is $\left|\partial_i\right|$. The total number of actions for node $i$ at time $k$ is 
$$1 + \left(\sum_{j=1}^{M}q_{i, k}^j\right)\left|\partial_i\right|,$$ where the ``$1$'' corresponds to staying silent. Hence, the probability of being silent is
$$\frac{1}{1 + \left(\sum_{j=1}^{M}q_{i, k}^j\right)\left|\partial_i\right|},$$ and the probability of transmitting the packet in $Q_{i, k}^\ell$ to neighbor $j$ is 
$$\frac{\setIn{q_{i, k}^\ell=1} \setIn{[\Xi_M]_{ij}=1}}{1 + \left(\sum_{j=1}^{M}q_{i, k}^j\right)\left|\partial_i\right|}.$$

\item {\it Adaptive age-based policies}: Nodes prefer transmitting cached packets with smaller age. Fix $\epsilon>0$. In time slot $k$, node $i$ stays silent with probability\footnote{We use $h_{i,k}^\ell+1$ instead of $h_{i,k}^\ell$ to avoid the case $h_{i, k}^\ell=0$.} $$\frac{e^{\epsilon}}{e^{\epsilon} + \sum_{\ell=1}^{M}\setIn{q_{i, k}^\ell=1} e^{1/(h_{i,k}^\ell+1)}},$$ and otherwise picks a packet $\ell$ with probability proportional to $e^{1/(h_{i,k}^\ell+1)}$, then chooses a receiver uniformly among its neighbors. Equivalently, the probability of transmitting the packet in $Q_{i, \ell}$ to neighbor $j$ with probability 
$$\frac{1}{d_i}\cdot\frac{\setIn{q_{i, k}^\ell=1} e^{1/(h_{i,k}^\ell+1)}}{e^{\epsilon} + \sum_{\ell=1}^{M}\setIn{q_{i, k}^\ell=1} e^{1/(h_{i,k}^\ell+1)}}.$$
\end{compactenum}

\subsection{Numerical Results}\label{sec: Numerical Results}

We now present the simulation results. First, we compare the performance of our algorithms with baselines on synthetic and real networks. Next, we examine transferability. Finally, we conduct a sensitivity analysis on the number of recurrent rounds.

\subsubsection{Performance on Synthetic and Real Networks}
\begin{figure*}[htbp]
\centering
\begin{subfigure}{0.32\linewidth}
    \centering
    \includegraphics[height=0.19\textheight]{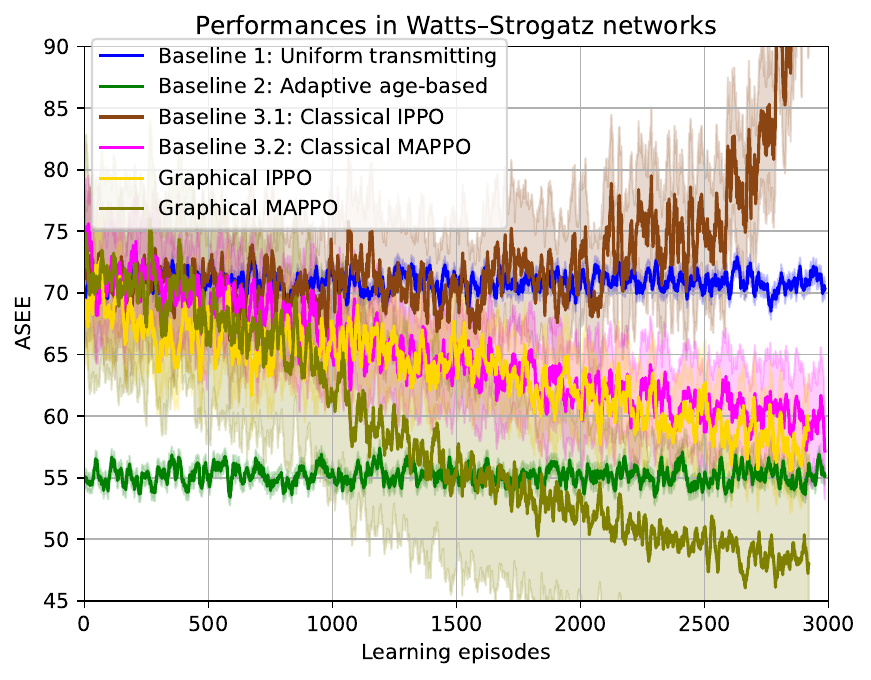}
    \caption{Watts–Strogatz networks}
\end{subfigure}
\begin{subfigure}{0.32\linewidth}
    \centering
    \includegraphics[height=0.19\textheight, trim=0 0 0 19,clip]{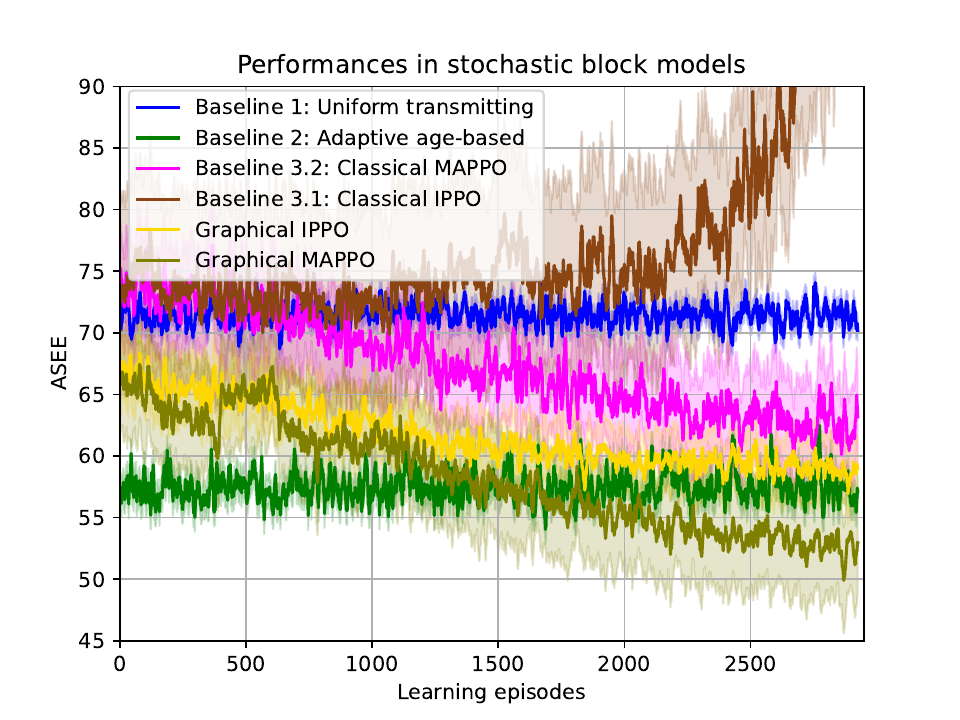}
    \caption{Stochastic block models}
\end{subfigure}
\begin{subfigure}{0.32\linewidth}
    \centering
    \includegraphics[height=0.19\textheight, trim=0 0 0 19,clip]{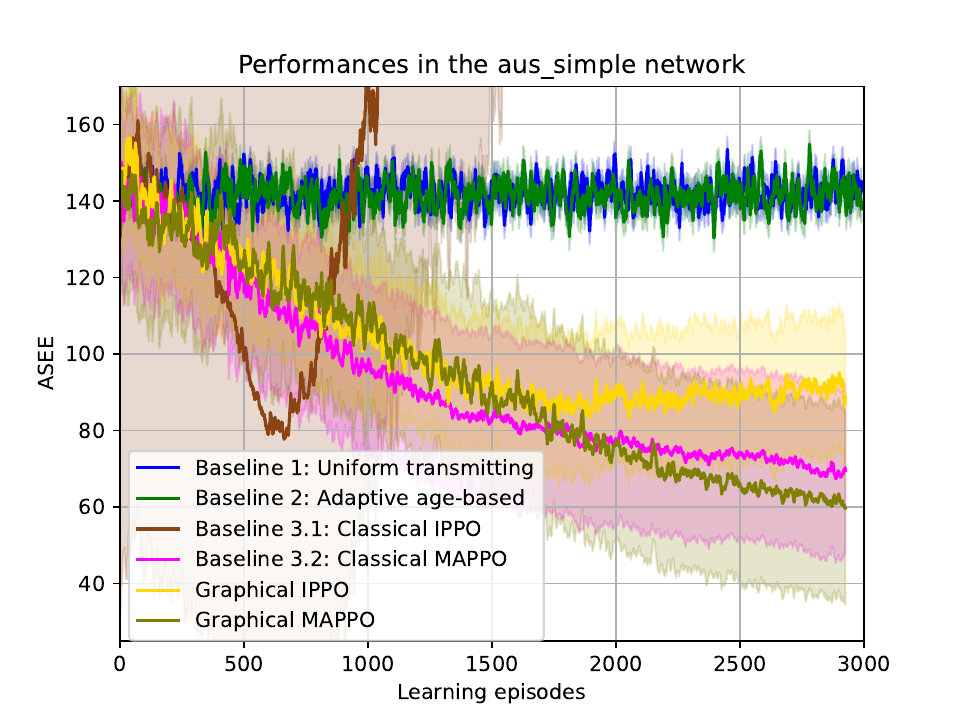}
    \caption{Real-data network ({\it aus\_simple})}
\end{subfigure}
\caption{Performance comparison between the proposed policies and baselines.}
\label{fig_performances}
\end{figure*}

The ASEE of our proposed policies and baselines in Watts–Strogatz graphs, stochastic block models are presented in Figs.~\ref{fig_performances}~(a) and (b).
 We derive the following insights: 
\begin{compactenum}[(i)]
\item Graphical IPPO policies outperform the classical IPPO policies, and the graphical MAPPO policies outperform the classical MAPPO policies, indicating the superiority of graphical MARL policies over classical MARL policies in our scenario.
\item Graphical MAPPO policies outperform graphical IPPO policies. This suggests that CTDE leads to better performance compared to independent learning in our setting.
\item For classical IPPO policies, the estimation error escalates with learning episodes due to the inherent non-stationarity of independent learning techniques. Comparing graphical IPPO policies with classical IPPO policies, we observe that graphical reinforcement learning exhibits greater resilience to non-stationarity.
\end{compactenum}

The ASEE on the real network is presented in Fig.\ref{fig_performances}(c), showing trends similar to those in Figs.\ref{fig_performances}(a) and (b). The advantage of graphical MAPPO is less pronounced than in Watts–Strogatz and stochastic block models, because although node indices are re-shuffled at the start of each episode, the graph structure remains fixed and the number of distinct permutations is limited. In relatively small networks, this reduces the benefit of graph-based learning. Nevertheless, the advantage of graphical MAPPO remains statistically significant: by $3000$ training episodes, the average ASEE gap compared to baselines exceeds $10$, confirming a meaningful performance gain.

\subsubsection{Transferability}
\begin{figure}[htbp]
	\centering
	\begin{subfigure}{0.86\linewidth}
		\centering
		\includegraphics[width=\linewidth]{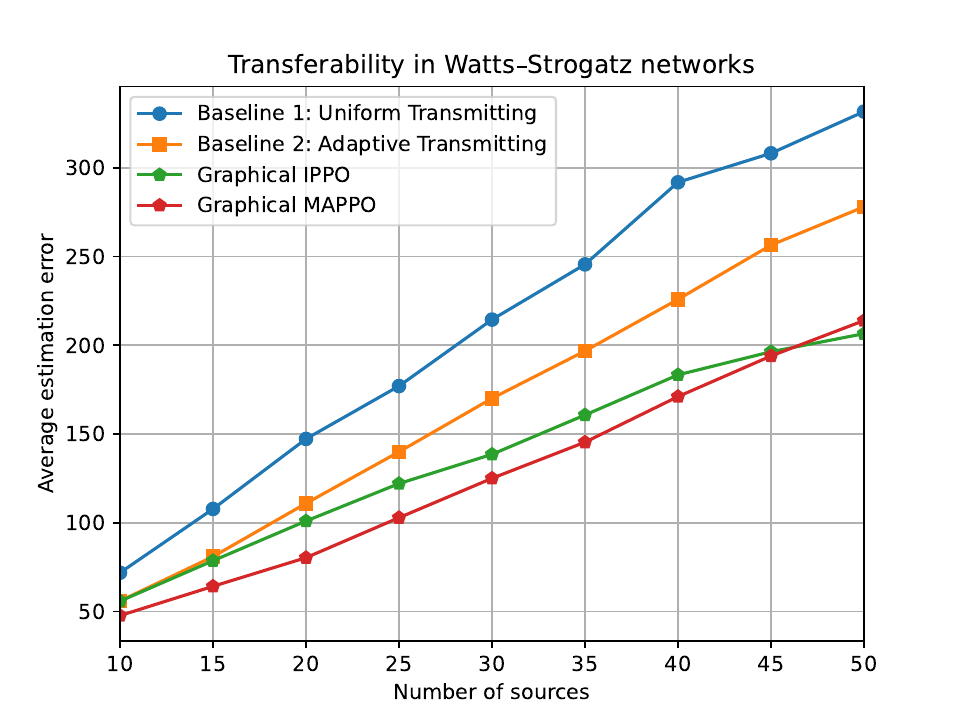}
		\caption{Transferability in Watts–Strogatz networks}
	\end{subfigure}
	\begin{subfigure}{0.86\linewidth}
		\centering
		\includegraphics[width=\linewidth]{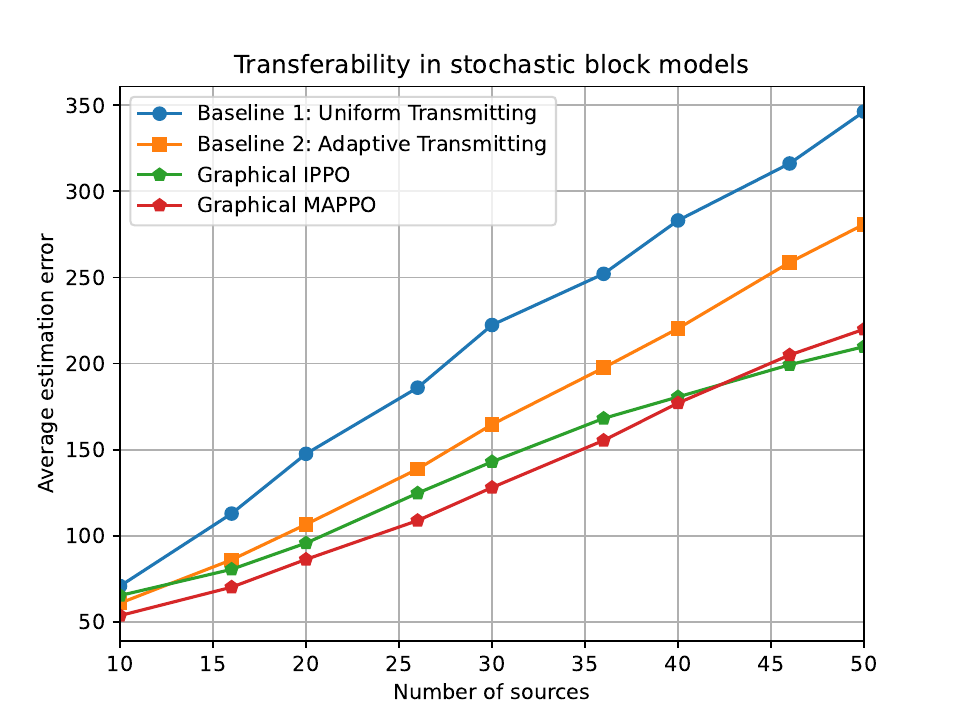}
		\caption{Transferability in stochastic block models}
	\end{subfigure}
	\caption{Transferability of proposed policies. The policies are trained on $10$-node networks and tested on networks with $M\in[10, 50]$ nodes.}
	\label{fig_transferability_1} 
\end{figure}

The transferability of our proposed frameworks is shown in Fig.~\ref{fig_transferability_1}. Models trained on $10$-node Watts–Strogatz networks and stochastic block models are applied to larger networks. As the number of nodes increases, the performance gap between our policies and the baselines widens. This indicates that the advantages observed in small networks persist in larger networks, with the growing gap further highlighting the amplified superiority of our policies.

Furthermore, the graphical IPPO policies exhibit better transferability than graphical MAPPO policies after reaching a certain network size ($M\approx45$ in Fig.~\ref{fig_transferability_1} (a) and $M\approx40$ in Fig.~\ref{fig_transferability_1} (b)). This is attributed to the fact that the transferability property holds only within the class of GNN architectures built on graph filters \cite{Transferability}. In contrast, in MAPPO, the critic GNN architectures are not built on graph filters. Therefore, this phenomenon occurs because the critic GNN architecture violates transferability. We believe that selecting critic GNN architectures built on graph filters would lead to graphical MAPPO policies outperforming graphical IPPO policies across all numbers of nodes.

\subsubsection{Sensitivity Analysis}
\begin{figure}
		\centering 
		\includegraphics[width=0.45\textwidth, height=0.26\textheight]{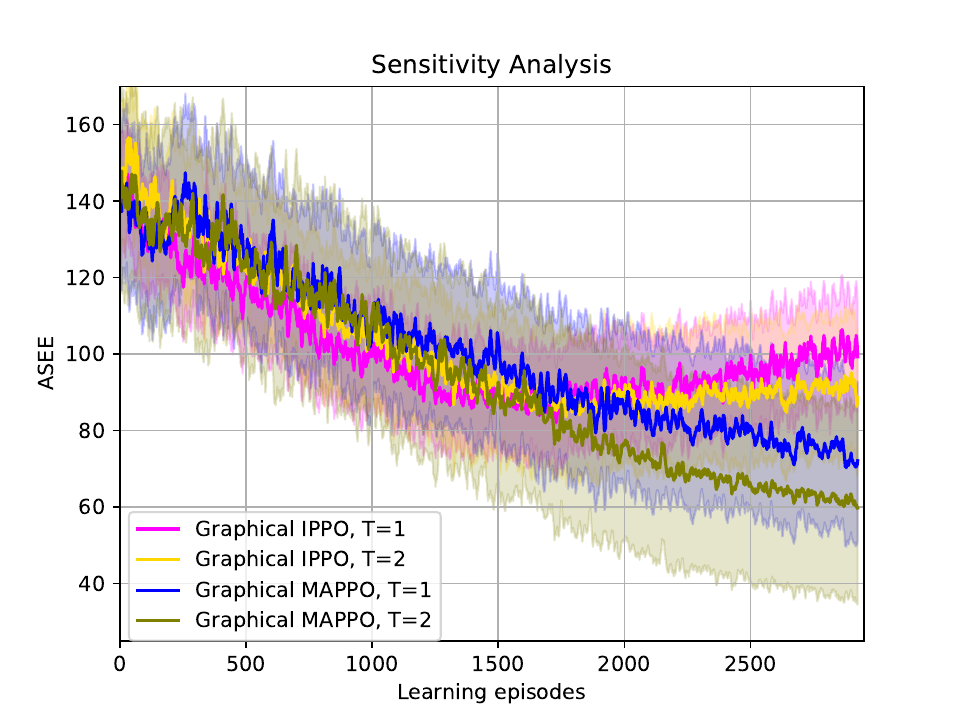} 
		\caption{Performances of proposed policies in the real network under different $T$.}
		\label{fig_sensitivity} 
	\end{figure}

	We are interested in the ASEEs of proposed policies with ($T=2$) and without ($T=1$) recurrence. In Fig.~\ref{fig_sensitivity}, we observe that in both graphical IPPO and graphical MAPPO policies, the ASEEs in policies with recurrence outperform those in policies without recurrence. This indicates that recurrence is beneficial in our proposed policies. Additionally, focusing on graphical IPPO, we observe that ASEEs in non-recurrent policies initially decrease before rising again, whereas ASEEs in recurrent policies also follow a decreasing-then-increasing trend but at a much slower rate. This suggests that recurrence enhances resilience to non-stationarity.

\section{Conclusion and Future Research}\label{sec: Conclusion}

In this paper, we study decentralized sampling and transmission policies for minimizing the time-average estimation error and/or the age of information in dynamic multi-hop wireless networks. We establish that, under oblivious policies, minimizing estimation error is equivalent to minimizing the age of information, providing a unifying perspective on these two objectives. 

Building on this insight, we develop a transferable graphical MARL framework for decentralized sampling and transmission. A key feature of the framework is its transferability, a learned policy can be directly deployed on dynamically evolving yet structurally similar graphs without re-training.  We further provide rigorous theoretical guarantees establishing this transferability. In contrast to prior studies on transferability in GNNs, our setting is fundamentally different, as we study a graphical MARL framework in which GNNs serve only as one component. Extensive simulations on both synthetic and real-world networks demonstrate consistent performance improvements over state-of-the-art baselines and strong positive transfer as the network scales.

Future work will focus on extending the proposed framework to more realistic communication environments, including noisy channels and partial observability, as well as improving training efficiency for very large-scale networks.

\bibliographystyle{IEEEtran}
\bibliography{references}

\clearpage
\appendices

\section{Proof of Theorem~\ref{thm: WRNN approximation on a generic graph}}\label{App: proof of transferability in GRNN}

Before proving the transferability in GRNN, we first prove the following  lemma. For clear representation, we denote
\begin{align*}
\Theta_1 =& (\Omega+\frac{\pi\kappa^\epsilon_{W_{\Xi_m}}}{\delta^\epsilon_{WW_{\Xi_m}}})\|W - W_{\Xi_m}\|, \\
\Theta_2 =& \Omega\epsilon + 2,\,\,\Theta_3 = 2\omega\epsilon.
\end{align*}

\begin{lemma}\label{lem: WNN approximation on a generic graph}
Let $T_{\mathcal{B}_i, W}$, $i\in\N$ be the WNNs, and assume that the convolutional filters that make up the
layers all satisfy Assumption~\ref{assu: filter Lipschitz}. Let $\rho_i$, $i\in\N$ satisfy Assumption~\ref{assu: activation function normalized Lipschitz}.   Define 
\begin{align*}
\left\{
\begin{aligned}
&E_1=T_{\mathcal{B}_1, W_{\Xi_m}}X_m,\\
&E_{i+1}=T_{\mathcal{B}_{i+1}, W_{\Xi_m}}E_{i}',\,\,E_{i}'=\rho_i(E_i),\,1\le i\le n-1,
\end{aligned}
\right.
\end{align*}
and 
\begin{align*}
\left\{
\begin{aligned}
&G_1=T_{\mathcal{B}_1, W}X,\\
&G_{i+1}=T_{\mathcal{B}_{i+1}, W}G_{i}',\,\,G_{i}'=\rho_i(G_i),\,1\le i\le n-1
\end{aligned}
\right.
\end{align*}
For any $0<\epsilon\le 1$, it holds that
\begin{compactenum}[(1)]
\item $\|E_n - G_n\|\le n (\Theta_1+\Theta_3)\|X\|+ \Theta_2\|X-X_m\|$;
\item $\|E'_n - G_n'\|\le n (\Theta_1+\Theta_3)\|X\|+\Theta_2\|X-X_m\|$.
\end{compactenum}
\end{lemma}
\begin{proof}
Part (1) is proved by mathematical induction, while part (2) follows immediately from part (1).

\noindent\textit{Proof of Part~(1).}

{\bf Step 1}. When $n=1$. $\|E_1 - G_1\| = \|T_{\mathcal{B}_1, W_{\Xi_m}}X_m - T_{\mathcal{B}_1, W}X\|$.  Since the convolutional filters that make up the layers of $T_{\mathcal{B}_1, \cdot}$ satisfy Assumption~\ref{assu: filter Lipschitz}, then \cite[Theorem~2]{Transferability},  for any $0<\epsilon\le 1$, it holds that
\begin{align*}
\|T_{\mathcal{B}_1, W_{\Xi_m}}X_m - T_{\mathcal{B}_1, W}X\|\le (\Theta_1+\Theta_3)\|X\| + \Theta_2\|X - X_m\|.
\end{align*}

{\bf Step 2}. We assume the inequality holds for all $k\leq n$, now we consider $k=n+1$. First, we expand the term $\|E_{n+1}-G_{n+1}\|$,
\begin{align*}
&\|E_{n+1}-G_{n+1}\| \\
& = \| T_{\mathcal{B}_{n+1}, W_{\Xi_m}}\rho_{n}(E_{n}) - T_{\mathcal{B}_{n+1}, W}\rho_{n}(G_{n})\|
\end{align*}
 By the triangle inequality, we have:
 \begin{align*}
 &\|E_{n+1}-G_{n+1}\| \\
 \le&\| T_{\mathcal{B}_{n+1},W_{\Xi_m}}\rho_{n}(E_n) - T_{\mathcal{B}_{n+1},W_{\Xi_m}}\rho_{n}(G_n)\big)\| \\
 +&\| T_{\mathcal{B}_{n+1},W_{\Xi_m}}\rho_{n}(G_{n}) - T_{\mathcal{B}_{n+1},W}\rho_{n}(G_{n})\|\\
 \triangleq&\D_1+\D_2.
 \end{align*} 

We first compute $\D_1$. Note that $T_{\mathcal{B}_{n+1},\cdot}$ satisfy Assumption~\ref{assu: filter Lipschitz},  the norm of the operator  $T_{\mathcal{B}_{n+1},\cdot}$ is bounded by $1$, hence
\begin{align*}
\D_1=&\| T_{\mathcal{B}_{n+1},W_{\Xi_m}}\rho_{n}(E_{n}) - T_{\mathcal{B}_{n+1},W_{\Xi_m}}\rho_{n}(G_{n})\| \\
=&\|T_{\mathcal{B}_{n+1}, W_{\Xi_m}}\big(\rho_{n}(E_{n}) - \rho_{n}(G_{n})\big)\|\\
\le & \|\rho_{n}(E_{n}) - \rho_{n}(G_{n})\|\\
\le &\|E_{n} - G_{n}||.
\end{align*}
The last inequality holds due to Assumption~\ref{assu: activation function normalized Lipschitz}.
Therefore, by assumption, 
\begin{equation}\label{eq: constantc}
\begin{aligned}
\D_1\le n (\Theta_1+\Theta_3)\|X\|+ \Theta_2\|X-X_m\|.
\end{aligned}
\end{equation}

Note that the activation function $\rho_{m}$ is pointwise non-linear. Then, for $\D_2$, again, by \cite[Theorem~2]{Transferability}, we have:
\begin{align*}
\D_2=&\|T_{\mathcal{B}_{n+1},W_{\Xi_m}}\rho_{n}(G_{n}) - T_{\mathcal{B}_{n+1},W}\rho_{n}(G_{n})||\\
\le&(\Theta_1+\Theta_3) ||\rho_{n}(G_{n})||. 
\end{align*}
Note that $T_{B_{n+1},W}$ satisfies Assumption~\ref{assu: filter Lipschitz} and $\rho_{n}$  satisfies Assumption~\ref{assu: activation function normalized Lipschitz}, so  
\begin{align*}
&\|\rho_{n}(G_{n})\|\le \|G_{n}\| = \|T_{\mathcal{B}_{n-1}, W}\rho_{n-1}(G_{n-1})\|\\
&\leq\|\rho_{n-1}(G_{n-1})\|\le \|G_{n-1}\|\cdots\leq \|G_1\| \\
&= \|T_{\mathcal{B}_1,W}X\|\le \|X\|.
\end{align*}
This implies that
\begin{align}\label{eq: constantb}
\D_2\le (\Theta_1+\Theta_3) \|X\| 
\end{align}
From \eqref{eq: constantc} and \eqref{eq: constantb}, we derive:
\begin{align*}
\D_1+\D_2\le (n+1) (\Theta_1+\Theta_3)\|X\|+ \Theta_2\|X-X_m\|.
\end{align*}
From {\bf Step 1} and {\bf Step 2}, we complete the proof.

\noindent\textit{Proof of Part (2).} 

Since $\rho_i$ with $i\in\N$ satisfy Assumption~\ref{assu: activation function normalized Lipschitz}, then 
\begin{align*}
\|E_n' - G_n'\|& = \|\rho_n(E_n) - \rho_n(G_n)\|\le \|E_n - G_n\|\\
&\le (n+1) (\Theta_1+\Theta_3)\|X\|+ \Theta_2\|X-X_m\|.
\end{align*}
\end{proof}

From the definition of WRNN in \eqref{eq: compact WRNN-1}, to prove Theorem~\ref{thm: WRNN approximation on a generic graph}, we only need to apply Lemma~\ref{lem: WNN approximation on a generic graph} repeatedly. The number of repetitions only depends on the number of recurrences $T$. Note that $\|X_t\|\le\eta_1$ and $\|X_t-X_{t,m}\|\le\eta_2$ for all $1\le t\le T$. After some algebra, we derive: 
\begin{align*}
\|Y-Y_m\|\le & \frac{T(1+T)}{2}(\Theta_1+\Theta_3)\eta_1 +  T \Theta_2\eta_2.
\end{align*}
This completes the proof.

\section{Proof of Theorem~\ref{thm: action distribution A}}\label{App: action distribution AA}
Let $T_{\cB}$, $T_{\cC}$, and $T_{\cD}$ be defined in \eqref{eq: compact WRNN-1}. Let $Y^{(j)}$ and $Y_m^{(j)}$ with $j\in\set{1,2}$ be defined in \eqref{eq:WRNNY1}--\eqref{eq:WRNNYn2}. We first prove the following lemma.  
\begin{lemma}\label{lem: action distributions}
Let $T_{\cB,W}$, $T_{\cC,W}$, and $T_{\cD,W}$ satisfy Assumption~\ref{assu: filter Lipschitz}, and let $\rho_1,\rho_2$ satisfy Assumption~\ref{assu: activation function normalized Lipschitz}. For any $0<\epsilon\le1$,  
\begin{align}\label{eq: action distribution}
&\left|\mathcal{X}(Y^{(1)}, Y^{(2)}) - \mathcal{X}_m(Y^{(1)}_m, Y^{(2)}_m)\right| \nonumber\\
&\le \|T_{W_\vartheta}\|\left(\|Y^{(2)}\|
+ \|Y_m^{(1)}\|\right)\eta_3.
\end{align}
\end{lemma}
\begin{proof}
From the construction $T_{W_\vartheta}$, it is continuous, hence is bounded \cite{Funtionalanalysis}, i.e.,
\begin{align*}
\|T_{W_\vartheta}x\|\le \|T_{W_\vartheta}\|\|x\|,\quad \forall x\in L^2.
\end{align*}
By definition, 
\begin{align*}
&\cX(Y^{(1)},Y^{(2)}) - \cX_m(Y_m^{(1)},Y_m^{(2)}) \\
&= \langle Y^{(1)},T_{W_\vartheta}Y^{(2)}\rangle - \langle Y_m^{(1)},T_{W_\vartheta}Y_m^{(2)}\rangle \\
&= \langle Y^{(1)}-Y_m^{(1)},\,T_{W_\vartheta}Y^{(2)}\rangle 
   + \langle Y_m^{(1)},\,T_{W_\vartheta}(Y^{(2)}-Y_m^{(2)})\rangle.
\end{align*}
Applying Cauchy–Schwarz and the operator bound gives
\begin{align*}
&\left|\cX(Y^{(1)},Y^{(2)}) - \cX_n(Y_m^{(1)},Y_m^{(2)})\right|\\
\le&\|T_{W_\vartheta}\|\left(\|Y^{(1)}-Y_m^{(1)}\|\|Y^{(2)}\|
+ \|Y_m^{(1)}\|\|Y^{(2)}-Y_m^{(2)}\|\right).
\end{align*}
From \eqref{eq:Delta}, $\|Y^{(j)}-Y_m^{(j)}\|\le \eta_3$ for $j\in\set{1, 2}$. Hence
\begin{align*}
&\left|\mathcal{X}(Y^{(1)}, Y^{(2)}) - \mathcal{X}_m(Y^{(1)}_m, Y^{(2)}_m)\right| \nonumber\\
&\le \|T_{W_\vartheta}\|\left(\|Y^{(2)}\|
+ \|Y_m^{(1)}\|\right)\eta_3.
\end{align*}
as claimed.
\end{proof}

The softmax function $F_{\text{softmax}}$ is Lipschitz \cite{SoftmaxLipschitz}, and its continuous extension $\tilde{F}_{\text{softmax}}$ is also Lipschitz \cite{OperatorLipschitz}. Based on the equivalence property of norms, there exists a constant $\Gamma$, independent of the WRNN $\Psi(\cdot)$, such that  
\begin{align}\label{eq: operator Lipschitz}
&\left|\tilde{F}_{\text{softmax}}\cX(Y^{(1)},Y^{(2)})- \tilde{F}_{\text{softmax}}\cX_m(Y^{(1)}_m,Y^{(2)}_m)\right| \nonumber\\
&\le \Gamma \left|\cX(Y^{(1)},Y^{(2)}) - \cX_m(Y^{(1)}_m,Y^{(2)}_m)\right|.
\end{align}
Substituting Lemma~\ref{lem: action distributions} into \eqref{eq: operator Lipschitz} gives
\begin{align*}
&\left|\tilde{F}_{\text{softmax}}\cX(Y^{(1)},Y^{(2)}) - \tilde{F}_{\text{softmax}}\cX_m(Y^{(1)}_m,Y^{(2)}_m)\right|\\
&\le \Gamma \|T_{W_\vartheta}\|\left(\|Y^{(2)}\|
+ \|Y_m^{(1)}\|\right)\eta_3.
\end{align*}

\end{document}